\documentclass[onecolumn]{IEEEtran}
\usepackage{amssymb,amsfonts,amsmath,amsthm}
\usepackage[dvips]{graphicx}
\usepackage{overpic}
\usepackage{subfigure}
\usepackage{color}
\usepackage{cases}
\usepackage{enumerate}
\usepackage{framed}
\definecolor{shadecolor}{rgb}{0.92,0.92,0.92}
\usepackage{listings}
\usepackage{graphicx}
\usepackage{float}
\usepackage{mathrsfs}
\usepackage[top=1.0in, bottom=1.08in, left=0.75in, right=0.75in]{geometry}
\usepackage{pgf}
\usepackage{tikz}
\usetikzlibrary{arrows, automata, positioning, calc, shapes}
\usepackage{comment}
\usepackage[numbers,sort&compress]{natbib}
\usepackage{caption}
\usepackage{hyperref}
\usepackage{bm} 
\usepackage[ruled,linesnumbered]{algorithm2e}
\usepackage{multirow}

\newtheorem{theorem}{Theorem}

\newtheorem{example}{Example}

\newtheorem{lemma}{Lemma}
\newtheorem{remark}{Remark}

\newtheorem{problem}{Problem}

\newtheorem{corollary}{Corollary}
\newtheorem{claim}{Claim}

\newcommand{\be}{\bm{e}}

\newcommand{\bg}{\bm{g}}
\newcommand{\bh}{\bm{h}}

\newcommand{\bu}{\bm{u}}
\newcommand{\bv}{\bm{v}}

\newcommand{\bx}{\bm{x}}
\newcommand{\by}{\bm{y}}

\newcommand{\cC}{\mathcal{C}}

\newcommand{\cH}{\mathcal{H}}

\newcommand{\cU}{\mathcal{U}}
\newcommand{\cV}{\mathcal{V}}
\newcommand{\cW}{\mathcal{W}}

\newcommand{\bbE}{\mathbb{E}}
\newcommand{\bbF}{\mathbb{F}}

\newcommand{\bbS}{\mathbb{S}}

\newcommand{\qbin}[2]{\genfrac{[}{]}{0pt}{}{#1}{#2}_q}

\usepackage{comment}
\allowdisplaybreaks

\title{Random Access in DNA Storage:\\ Algorithms, Constructions, and Bounds}
\author{Chen Wang\! and Eitan Yaakobi\!
\thanks{The research of Chen Wang was supported in part at the Technion by a fellowship from the Lady Davis Foundation. The work of Eitan Yaakobi was partially funded by the European Union (DiDAX, 101115134). Views and opinions expressed are however those of the authors only and do not necessarily reflect those of the European Union or the European Research Council Executive Agency. Neither the European Union nor the granting authority can be held responsible for them. This research was also partially supported in part by the Israel Science Foundation (ISF) Grant 2462/24. This work was also supported by a DFG Middle-East Grant, under grant number 554629494. }
\thanks{Chen~Wang and Eitan~Yaakobi are with the Department of Computer Science, Technion --- Israel Institute of Technology, Haifa 3200003, Israel (e-mails:\href{mailto:cwang@campus.technion.ac.il}{cwang@campus.technion.ac.il}, \href{mailto:yaakobi@cs.technion.ac.il}{yaakobi@cs.technion.ac.il})}
\thanks{An earlier version of this paper has been accepted for presentation at the 2026 IEEE International Symposium on Information Theory (ISIT), Guangzhou, China, June 28--July 3, 2026~\cite{wang2026randomisit}.}
}
\date{\today}

\begin{document}
	\maketitle
	\begin{abstract}
		As DNA data storage advances toward practical deployment, minimizing sequencing coverage depth is critical for reducing both operational costs and retrieval latency. In this work, we address the recently studied \emph{random access problem}, which concerns the challenge of recovering a specific information strand from a DNA-based storage system. In this setting, $k$ information strands are encoded into $n$ strands using a generator matrix $G$, and each sequencing read returns one encoded strand sampled uniformly at random with replacement from the $n$ strands. The objective is twofold: to minimize the expected number of samples required to recover an arbitrary information strand in the random access setting, and to minimize the expected number of samples required to recover all information strands in the non-random access setting. For the random access setting, we propose a novel formula for the exact expected number of samples required to recover a specific information strand, which yields an algorithm that computes the exact expected number of samples with a time complexity of $O(n)$ for fixed field size $q$ and dimension $k$. Moreover, we derive explicit formulas for the average and maximum expected number of samples, facilitating an efficient search for optimal generator matrices for small parameters. In addition to theoretical analysis, we present new constructions that improve the best-known upper bounds from $0.8815k$ to $0.8811k$ for $k=3$, and from $0.8637k$ to $0.8629k$ for $k=4$, for sufficiently large $q$. We also establish a tighter theoretical lower bound on the expected number of samples, improving upon the current state-of-the-art bound. Notably, this bound confirms the optimality of the simple parity code for the case $n = k+1$ over any field size $q$. Finally, for the non-random access setting to recover all information strands, we propose new lower bounds and constructions, which characterize the asymptotic behavior of the expected number of samples required to recover all information strands.
	\end{abstract}

	\section{Introduction}
	With the rapid advancement of synthesis and sequencing technologies, DNA molecules have emerged as a promising medium for long-term digital information storage, primarily owing to their exceptional density and durability \cite{church2012next,goldman2013towards}. In a typical DNA storage pipeline, the data is first encoded into sequences over the DNA alphabet $\{A,T,C,G\}$ and then synthesized into artificial DNA strands. These strands are stored in a container, where each strand exists in numerous copies, but in a completely unordered manner. When a user wishes to recover the stored data, it is necessary to sequence the strands to obtain a multiset of reads. Once enough reads have been obtained, the user is able to reconstruct the original data. The sequencing process, performed using a DNA sequencer, remains one of the primary operational bottlenecks in any DNA storage system \cite{anavy2019data, erlich2017dna, organick2018random, yazdi2017portable}. Although numerous studies have demonstrated the tremendous potential of DNA as a data storage medium \cite{bar2025scalable,blawat2016forward,bornholt2016dna,tabatabaei2015rewritable,banal2021random}, its viability as a practical alternative to existing storage technologies remains limited by challenges related to cost and efficiency \cite{shomorony2022information, yazdi2016dna}. A primary factor underlying these challenges is the required \emph{coverage depth}, defined as the number of reads required for successful data retrieval. Coverage depth directly impacts both system latency and cost, underscoring the critical need for optimization \cite{chandak2019improved, erlich2017dna} and motivating the study of strategies to reduce coverage depth in DNA storage systems.
	
	In this work, we focus on a specific aspect of coverage depth optimization, known as the \emph{random access problem}, initially introduced by Bar-Lev $\textit{et al.}$ \cite{daniella2023cover}. In this framework, it is assumed that $k$ information strands, which encode the user data, are encoded into $n$ strands using a generator matrix $G \in \mathbb{F}_q^{k \times n}$. During the retrieval process, the encoded strands are sampled uniformly at random with replacement, which is mathematically equivalent to sampling the columns of $G$. Bar-Lev $\textit{et al.}$ considered two variants of this problem: the \emph{random access} and the \emph{non-random access} settings. In the former, the objective is to recover a single specific information strand from among the $k$ information strands, while in the latter, the goal is to recover all $k$ information strands. In both scenarios, the aim is to minimize the expected number of samples required to recover the desired information strand(s).

	For the random access setting, Bar-Lev $\textit{et al.}$ \cite{daniella2023cover} established an upper bound of $0.9595k$ on the expected number of samples for any field size $q$ and dimension $k$. Additionally, they derived two lower bounds: one of $\frac{k + 1}{2}$ and another of $n - \frac{n(n - k)}{k}(H_n - H_{n - k})$, both for the maximum expected number of samples required to recover any information strand. Following the work of Bar-Lev $\textit{et al.}$ \cite{daniella2023cover}, the random access setting has been widely studied \cite{gruica2024combinatorial,gruica2024geometry,boruchovsky2025making,bodur2025random,bertuzzo2025coverage,abraham2024covering,tan2025labeled,barak2025conjectures,grunbaum2025general}, with most efforts focused on constructing generator matrices that achieve a lower expected number of samples. Recently, Gruica $\textit{et al.}$ \cite{gruica2024combinatorial} introduced new techniques to investigate the random access problem, effectively capturing its combinatorial nature and identifying structural properties of generator matrices that are advantageous for retrieval. Furthermore, Boruchovsky $\textit{et al.}$ \cite{boruchovsky2025making} determined the optimal construction for the case $k = 2$ and proposed constructions for larger $k$, while Bodur $\textit{et al.}$ \cite{bodur2025random} further improved upon these constructions. Although significant progress has been made, several critical problems remain unsolved. Efficiently computing the expected number of samples remains challenging, especially for large $n$, and the optimal constructions for $k \geq 3$ are still unknown.

	For the non-random access setting, Bar-Lev $\textit{et al.}$ \cite{daniella2023cover} established a connection to the classical \emph{Coupon Collector's Problem} \cite{erdHos1961classical,feller1991introduction,flajolet1992birthday,newman1960double}. They proved that the expected number of samples required to recover all information strands is at least $n(H_n - H_{n-k})$, where $H_i$ denotes the $i$-th harmonic number. Furthermore, they showed that this bound is tight for maximum distance separable (MDS) codes, and that no coding scheme can achieve a smaller expectation. Consequently, MDS codes are optimal for minimizing the expected number of samples when the objective is to recover the entire dataset. The non-random access setting was subsequently extended to incorporate composite DNA letters \cite{cohen2025optimizing,anavy2019data}, and was further investigated in \cite{preuss2024sequencing,sokolovskii2024coding} in the context of combinatorial composites of DNA shortmers \cite{preuss2024efficient}. A related line of work was presented in \cite{chandak2019improved}, where the authors analyzed the trade-offs between reading costs, which are directly determined by coverage depth, and writing costs. As the existence of MDS codes typically requires a large field size, Bertuzzo $\textit{et al.}$ \cite{bertuzzo2025coverage} further studied this problem for small fields. They established a new formula for the expected number of samples required to recover all information strands, and investigated some specific constructions over small fields. 

	In this work, we investigate both the random access and non-random access settings. For the random access setting, we propose a novel formula for the exact expected number of samples required to recover a specific information strand, which leads to an efficient algorithm that computes the exact expected number of samples with a time complexity of $O(n)$ for fixed field size $q$ and dimension $k$. Moreover, we derive explicit formulas for the average and maximum expected number of samples, which enable an efficient search for optimal generator matrices for small parameters. For sufficiently large code length $n$, we present new constructions that improve the best-known upper bounds from $0.8815k$ to $0.8811k$ for $k=3$, and from $0.8637k$ to $0.8629k$ for $k=4$, for sufficiently large $q$. To narrow the gap between constructions and lower bounds, we establish a tighter theoretical lower bound on the average expected number of samples, improving upon the current state-of-the-art bounds. In particular, our bound demonstrates the optimality of the simple parity code for the case $n = k+1$ over any field size $q$. For the non-random access setting, we propose new lower bounds that take the field size $q$ into account, and provide corresponding constructions. These lower bounds and constructions characterize the asymptotic behavior of the expected number of samples required to recover all information strands.

	The rest of the paper is organized as follows. In Section \ref{sec:pre}, we introduce the necessary preliminaries and formally define the problems studied in this work. Section \ref{sec:algorithm} focuses on the random access setting, where we present a new formula and our algorithm for computing the expected number of samples. In Section \ref{sec:upper bounds}, we derive explicit formulas for the average and maximum expected number of samples, and present new constructions that improve the best-known upper bounds. Section \ref{sec:lower bounds} is dedicated to establishing new lower bounds on the expected number of samples. In Section \ref{sec:expectation_number_of_all_samples}, we investigate the non-random access setting, providing new lower bounds and constructions for recovering all information strands. Finally, we conclude the paper in Section \ref{sec:conclusion} and discuss potential directions for future research.

	\section{Preliminaries}\label{sec:pre}
	\subsection{Notations and Problem Formulation}
	For any integers $m, n$, let $[m, n]$ denote the set $\{m, m+1, \ldots, n\}$ and $[n] := [1, n]$. Let $H_n$ denote the $n$-th harmonic number, i.e., $H_n := 1 + \frac{1}{2} + \cdots + \frac{1}{n}$. Let $q$ be a prime power and $\mathbb{F}_q$ denote the finite field with $q$ elements. We denote by $\qbin{\cdot}{\cdot}$ the Gaussian binomial coefficient. Let $\mathbb{F}_q^k$ be the $k$-dimensional vector space over $\mathbb{F}_q$. For $i \in [k]$, let $\be_i = (0, 0, \ldots, 1, \ldots, 0) \in \mathbb{F}_q^k$ denote the $i$-th standard basis vector, with $1$ in the $i$-th position and $0$ elsewhere. For a vector $\bu \in \mathbb{F}_q^k$, let $\mathrm{wt}(\bu)$ denote its Hamming weight, i.e., the number of nonzero coordinates. For an integer $\ell \geq 0$ and vectors $\bg_1, \bg_2, \ldots, \bg_{\ell} \in \mathbb{F}_q^k$, let $\langle \bg_1, \bg_2, \ldots, \bg_{\ell} \rangle$ denote their $\mathbb{F}_q$-linear span. Finally, let $\bbS_{q, k}^t$ denote the set of all $t$-dimensional subspaces of $\mathbb{F}_q^k$.

	For a space $\cU$, let $\mathrm{dim}(\cU)$ denote its dimension. For two spaces $\cU$, $\cV$, we write $\cV \leq \cU$ if $\cV$ is a subspace of $\cU$. A subspace $\cU \in \bbS_{q, k}^t$ is called \emph{standard} if $\cU = \langle \be_{i_1}, \ldots, \be_{i_t} \rangle$ for some standard basis vectors $\be_{i_1}, \ldots, \be_{i_t}$. For a vector $\bu$, let $\mathrm{supp}(\bu)$ denote its support. That is, $\mathrm{supp}(\bu) := \{\be_i : i \in [k],\, u_i \neq 0\}$. For a space $\cU \leq \mathbb{F}_q^k$, define its support as $\mathrm{supp}(\cU) = \bigcup_{\bu \in \cU} \mathrm{supp}(\bu)$.

	Let $\Sigma = \{A, T, C, G\}$ be the DNA alphabet. In DNA storage systems, data is represented as a length-$k$ information vector $\bx = (x_1, \ldots, x_k)$, where each $x_i \in \Sigma^{\ell}$ corresponds to an information strand. To enable linear encoding and decoding, we embed $\Sigma^{\ell}$ into a finite field $\mathbb{F}_q$ and encode $\bx$ into a codeword $\by = (y_1, \ldots, y_n) \in \mathbb{F}_q^n$ using a rank-$k$ generator matrix $G$. Although embedding $\Sigma^{\ell}$ into $\mathbb{F}_q$ strictly requires that $4^{\ell} \leq q$, in this work we consider arbitrary prime powers $q$ for generality.
	
	In the random access setup, the information vector $\bx$ is encoded using a generator matrix $G = [\bg_1, \ldots, \bg_n] \in \mathbb{F}_q^{k \times n}$, and the user can recover the information strands by sampling each coordinate of $\by = \bx G$ uniformly at random. In the random access setting, the user aims to recover a single specific information strand, whereas in the non-random access setting, the goal is to recover all information strands. In both cases, the objective is to minimize the expected number of samples required to recover the desired information strands. The following example illustrates the model.

	\begin{example}\label{exa:1}
	Let $q = 2$, $k = 2$, and $n = 5$. Consider $\bx = (x_1, x_2)$ and a generator matrix $$G = \begin{bmatrix}
			1, 0, 1, 0, 1\\
			0, 1, 0, 1, 1\end{bmatrix}\in\bbF_2^{2\times 5}$$
	which encodes the information as $(x_1, x_2)G = (x_1, x_2, x_1, x_2, x_1 + x_2) \in \mathbb{F}_2^5$. To recover $x_i$, the user samples a coordinate of $\bx G$ uniformly at random. That is, in each sample, the probabilities of sampling $x_1$, $x_2$, or $x_1 + x_2$ are $\frac{2}{5}$, $\frac{2}{5}$, and $\frac{1}{5}$, respectively. One can verify \cite{boruchovsky2025making} that the expected number of samples required to recover either information strand is approximately $1.917 < 2$. When the user aims to recover both $x_1$ and $x_2$, the expected number of samples required is $\frac{31}{12}$, as the formula established in Lemma \ref{lem:whole_alpha}.
	\end{example}

	As illustrated in Example \ref{exa:1}, for any set of $k$ information strands encoded by a generator matrix $G$, each coded strand corresponds to a column of $G$. Based on this observation, we now formally introduce the model considered in this work. Let $G = [\bg_1, \ldots, \bg_n] \in \mathbb{F}_q^{k \times n}$ be a rank-$k$ matrix. In the random access model, we sample each column of $G$ uniformly at random. For $i \in [k]$, let $\tau_i(G)$ be the random variable counting the minimum number of columns of $G$ that must be sampled with replacement until the standard basis vector $\be_i$ lies in their $\mathbb{F}_q$-span. Similarly, let $\tau(G)$ be the random variable counting the minimum number of columns of $G$ that must be sampled with replacement until the entire space $\mathbb{F}_q^k$ is spanned. Define $T_{\max}(G) = \max_{i \in [k]} \mathbb{E}[\tau_i(G)]$ as the maximum expected number of samples and $T_{\mathrm{ave}}(G) = \frac{1}{k} \sum_{i = 1}^k \mathbb{E}[\tau_i(G)]$ as the average expected number of samples. Notably, recovering the $i$-th standard basis vector $\be_i$ from the columns of $G$ is equivalent to recovering the $i$-th information strand in the code generated by $G$. It is important to note that all expectations in our analysis depend on the generator matrix $G$. To study the optimal performance under the random access setting, we define the optimal average and maximum expected number of samples as follows:
	\begin{align*}
	T_{\mathrm{ave}}(q, n, k) := \min_{G \in \mathbb{F}_q^{k \times n}} T_{\mathrm{ave}}(G), \
	T_{\max}(q, n, k) := \min_{G \in \mathbb{F}_q^{k \times n}} T_{\max}(G),
	\end{align*}
	along with their asymptotic behavior as $n$ approaches infinity:
	\begin{align*}
	T_{\mathrm{ave}}(q, k) := \liminf_{n \to \infty} T_{\mathrm{ave}}(q, n, k), \
	T_{\max}(q, k) := \liminf_{n \to \infty} T_{\max}(q, n, k).
	\end{align*}
	For the non-random access setting, let $$T(q, n, k) := \min_{G\in\bbF_q^{k\times n}}\bbE[\tau(G)]$$ denote the minimum expected number of samples required to recover all information strands for given parameters $q$, $n$, and $k$. For any generator matrix $G$, let
	\[
	w_{\bu}(G) := \frac{|\{j\in[n] : \bg_j=\bu\}|}{n}
	\]
	be the fraction of columns equal to $\bu$. Equivalently, $w_{\bu}(G)$ is the probability of sampling $\bu$ under $G$. Hence, $\bbE[\tau_i(G)]$ and $\bbE[\tau(G)]$ depend only on the column distribution $(w_{\bu}(G))_{\bu\in\bbF_q^k}$. In this work, we formally introduce the following problems. For the random access setting, we first consider the problem of determining $\mathbb{E}[\tau_i(G)]$ for a given generator matrix $G$.
	\begin{problem}\label{pro:algorithm}
		For fixed $k$, $q$, and a generator matrix $G \in \mathbb{F}_q^{k \times n}$, determine the relationship between $\mathbb{E}[\tau_i(G)]$ and the column distribution $(w_{\bu}(G))_{\bu\in\bbF_q^k}$. Furthermore, develop a linear-time algorithm to determine $\mathbb{E}[\tau_i(G)]$ for each $i \in [k]$ for fixed $q$ and $k$.
	\end{problem}

	We also consider the problem of determining the minimum average and maximum expected number of samples.
	\begin{problem}\label{pro:random_access}
		Given $q$, $k$, and $n$ with $n \geq k$, establish bounds on the smallest possible average and maximum expected number of samples, denoted by $T_{\mathrm{ave}}(q, n, k)$ and $T_{\max}(q, n, k)$ respectively, and characterize their asymptotic behavior $T_{\mathrm{ave}}(q, k)$ and $T_{\max}(q, k)$ as $n \to \infty$.
	\end{problem}

	For the non-random access setting, we consider the problem of determining the minimum expected number of samples required to recover all information strands.

	\begin{problem}\label{pro:whole_access}
		Given $q$, $k$, and $n$ with $n \geq k$, establish bounds on the smallest possible expected number of samples required to recover all information strands $T(q, n, k)$.
	\end{problem}
	
	\subsection{Related Works}
	The random access model has been extensively studied in \cite{gruica2024combinatorial,gruica2024geometry,boruchovsky2025making,daniella2023cover,bodur2025random,bertuzzo2025coverage,abraham2024covering,tan2025labeled,barak2025conjectures,grunbaum2025general}. A fundamental problem in the random access setting is to determine $\mathbb{E}[\tau_i(G)]$ for a given generator matrix $G$. In \cite{gruica2024combinatorial}, the authors introduced Lemma \ref{lem:alpha} to compute $\mathbb{E}[\tau_i(G)]$. This requires defining $\alpha_i^s(G)$ as the number of subsets $S \subseteq [n]$ of size $s$ such that the standard basis vector $\be_i$ lies in the $\mathbb{F}_q$-span of the columns indexed by $S$, i.e., $$\alpha_i^s(G) = |\{S\subseteq [n]:|S| = s, \be_i\in\langle\bg_j:j\in S\rangle\}|.$$
	\begin{lemma}[See \cite{gruica2024combinatorial}, Lemma 1]\label{lem:alpha}
		For every $i\in [k]$, the expected value of $\tau_i(G)$ is given by: $$\bbE[\tau_i(G)] = nH_n - \sum_{s = 1}^{n - 1}\frac{\alpha_i^s(G)}{\binom{n-1}{s}}.$$
	\end{lemma}
	To the best of our knowledge, computing $\alpha_i^s(G)$ for a general matrix $G$ remains a computationally challenging task, even for fixed $k$ and $q$. This computation requires identifying all subsets of columns of size $s$ whose span contains the basis vector $\be_i$. For a given $s$, evaluating $\alpha_i^s(G)$ would necessitate examining all $\binom{n}{s}$ subsets. For each subset of size $s$, verifying whether $\be_i$ lies in the span of its columns requires $O(k s \min\{k, s\})$ operations using Gaussian elimination. Consequently, the worst-case time complexity of this approach is $O(k^2 n 2^n)$.
	
	Another significant problem is constructing the optimal generator matrix $G$ that minimizes the maximum or average expected number of samples. Bar-Lev $\textit{et al.}$ \cite{daniella2023cover} showed that for a systematic generator matrix $G$ of an MDS code (i.e., every set of $k$ columns of $G$ is linearly independent), we have $T_{\max}(G) = k$. Gruica $\textit{et al.}$ \cite{gruica2024combinatorial} further generalized this result, demonstrating that for any recovery-balanced code, $T_{\max}(G) = k$. Notably, many practical codes satisfy the recovery-balanced property (see \cite[Section 5]{gruica2024combinatorial} for details).

	To achieve a lower expected number of samples, Bar-Lev $\textit{et al.}$ \cite{daniella2023cover} constructed a matrix $G$ and showed that $T_{\mathrm{ave}}(G) \leq 0.9595k$ for all field sizes $q \geq 2$, and conjectured that $T_{\mathrm{ave}}(G) \leq 0.9456k$. This conjecture was subsequently proved in \cite{gruica2024geometry}, which also presented constructions achieving $T_{\mathrm{ave}}(G) \leq 0.88\overline{22}k$ for $k = 3$ and sufficiently large $q$. Boruchovsky $\textit{et al.}$ \cite{boruchovsky2025making} determined the exact expression for $T_{\max}(q, k = 2)$, showing that	
	$$T_{\max}(q, k = 2) = 1 + \frac{2q^2 - q(\sqrt{2} + 1) - 2 + \sqrt{2}}{q^2(1 + \sqrt{2}) - q(2 + \sqrt{2})},$$
	which yields $\liminf_{q \to \infty} T_{\max}(q, k = 2) \approx 0.914 \cdot 2$. They also established upper bounds $\liminf_{q \to \infty} T_{\max}(q, k = 3) \leq 0.881\overline{66} \cdot 3$ and $\liminf_{q \to \infty} T_{\max}(q, k = 4) \leq 0.86375 \cdot 4$, along with a construction that works for all $k$. More recently, Bodur $\textit{et al.}$ \cite{bodur2025random} improved the construction for $k = 3$, showing that $\liminf_{q \to \infty} T_{\max}(q, k = 3) \lesssim 0.881542 \cdot 3$. Additionally, they determined that the optimal construction over $\mathbb{F}_q$ satisfies $T_{\max}(q, n, k = 3) \leq \frac{8\sqrt{3}\pi - 18}{27}$ for any $q \geq 2$. Nevertheless, the optimal constructions for $k \geq 3$ remain unknown.
	
    For the non-random access setting, the expected number of samples $\mathbb{E}[\tau(G)]$ required to recover all information strands has been studied in \cite{daniella2023cover,bertuzzo2025coverage}. Bar-Lev $\textit{et al.}$ \cite{daniella2023cover} showed that for any generator matrix $G\in\mathbb{F}_q^{k\times n}$,
    $$\mathbb{E}[\tau(G)]\geq n(H_n - H_{n-k}),$$
    with equality holding when $G$ is a generator matrix of an MDS code. As the existence of MDS codes typically requires a large field size, Bertuzzo $\textit{et al.}$ \cite{bertuzzo2025coverage} extended this analysis to small field sizes. Let $\alpha^s(G)$ denote the number of $s$-element column subsets $S \subseteq [n]$ whose $\mathbb{F}_q$-span equals $\mathbb{F}_q^k$, i.e.,
    $$\alpha^s(G) = \left|\left\{S\subseteq [n]:|S| = s, \left\langle\bg_j:j\in S\right\rangle = \mathbb{F}_q^k\right\}\right|.$$
    They provided the following result to compute $\mathbb{E}[\tau(G)]$.
	\begin{lemma} [See \cite{bertuzzo2025coverage}, Proposition 2]\label{lem:whole_alpha}
		Let $G\in\bbF_q^{k\times n}$ be a generator matrix, it holds that
		\[
			\bbE[\tau(G)] = nH_n - \sum_{s = k}^{n - 1}\frac{\alpha^s(G)}{\binom{n-1}{s}}.
		\]
	\end{lemma}
	They also analyzed the expected number of samples for specific code constructions, which are summarized as follows.
    \begin{itemize}
        \item Let $G$ be the generator matrix of the $q$-ary simplex code $\cC\subset \bbF_q^n$ of dimension $k$, where $n = (q^{k} - 1)/(q - 1)$. Then, it holds that 
		\[
			\bbE[\tau(G)] = k + \sum_{i = 1}^k\frac{q^{i - 1} - 1}{q^{k} - q^{i - 1}}.
		\]
        \item Let $G$ be the generator matrix of the $q$-ary Hamming code $\cC\subset \bbF_q^n$ of dimension $k$ and redundancy $r$, where $n = (q^{r} - 1)/(q - 1)$ and $k = n - r$. Then, it holds that 
		\[
			\bbE[\tau(G)] = n H_n-\sum_{\ell=1}^r \frac{1}{\binom{n-1}{n-\ell}} \frac{\prod_{i=0}^{\ell-1} \frac{q^r-q^i}{q-1}}{\ell!}.
		\]
    \end{itemize}
	However, the lower bound remains relatively weak for small field sizes $q$, and optimal constructions achieving lower expected sample numbers remain elusive, particularly for small $q$ and in regimes where $n$ does not grow exponentially in $k$.

    In this work, our contributions are as follows:
    \begin{itemize}
        \item We develop a novel formula for the expected number of samples, which yields an efficient algorithm to compute it for a given generator matrix $G \in \mathbb{F}_q^{k \times n}$, achieving a time complexity of $O(n)$ for fixed field size $q$ and information length $k$.
        \item We derive explicit formulas for the average and maximum expected number of samples, enabling an efficient search for optimal generator matrices for small parameters. Furthermore, we present new code constructions that improve the best-known upper bound from $0.8815k$ to $0.8811k$ for $k=3$, and achieve an upper bound of $0.8629k$ for $k=4$, for sufficiently large $n$ and $q$.
        \item We establish a tighter theoretical lower bound on the expected number of samples $\mathbb{E}[\tau_i(G)]$, improving upon the current state-of-the-art bounds. In particular, this bound confirms the optimality of the simple parity code for the case $n = k+1$ over any field size $q$.
        \item We study the expected number of samples $\mathbb{E}[\tau(G)]$ required to recover all information strands, providing new lower bounds and constructions. Our results improve upon state-of-the-art bounds for small field sizes $q$ and small $n$, and characterize the asymptotic order of $\mathbb{E}[\tau(G)]$.
    \end{itemize}
	
	\section{Efficient Algorithm to Determine the Expected Number of Samples}\label{sec:algorithm}
	
	In this section, we develop an efficient algorithm to compute the expected number of samples $\mathbb{E}[\tau_i(G)]$ for a given generator matrix $G \in \mathbb{F}_q^{k \times n}$. For any subspace $\cU \leq \mathbb{F}_q^k$, define $w_{\cU}(G) = \sum_{\bu \in \cU} w_{\bu}(G)$ as the total proportion of columns that lie in $\cU$. A set of columns of $G$ that fails to recover $\be_i$ is called a \emph{bad set for $\be_i$}. Let $\gamma_i^s(G) = \binom{n}{s} - \alpha_i^s(G)$ denote the number of bad sets for $\be_i$ of size $s$. Lemma \ref{lem:formula} establishes an explicit relationship between $\mathbb{E}[\tau_i(G)]$ and the proportions $w_{\cU}(G)$ for subspaces $\cU$ that do not contain $\be_i$.
	
	\begin{lemma}\label{lem:formula}
		For any generator matrix $G = [\bg_1, \ldots, \bg_n]\in\bbF_q^{k\times n}$, it holds that
		\begin{align*}
			\bbE[\tau_i(G)] = 1 + \sum_{t = 1}^{k - 1}\sum_{\be_i\notin\cU\in\bbS_{q, k}^t}\sum_{r = t}^{k - 1}(-1)^{r- t}q^{\binom{r- t}{2}}\left(\qbin{k - t}{r - t} - \qbin{k - t - 1}{r - t - 1}\right)\frac{w_{\cU}(G)}{1 - w_{\cU}(G)}.
		\end{align*}
	\end{lemma}

	Before proving Lemma~\ref{lem:formula}, we introduce the following well-known combinatorial identity.
	\begin{claim}\label{cla:combinatorial formula}
		For any integer $n\geq 1$ and $0\leq r\leq n$, it holds that 
		$$\sum_{i = 0}^{n}\frac{\binom{r}{i}}{\binom{n}{i}} = \frac{n + 1}{n - r + 1}.$$
	\end{claim}
	
	\begin{IEEEproof}[Proof of Claim \ref{cla:combinatorial formula}]
		Observe that for $0\leq i\leq n$, we have 
		\[
			\frac{\binom{r}{i}}{\binom{n}{i}} = \frac{r!}{i!(r - i)!}\cdot \frac{i!(n - i)!}{n!} = \frac{r!(n-i)!}{n!(r-i)!}=\frac{1}{\binom{n}{r}}\binom{n - i}{n - r}.
		\]
		Therefore, we have
		\begin{align*}
			\sum_{i = 0}^{n}\frac{\binom{r}{i}}{\binom{n}{i}} = \frac{1}{\binom{n}{r}}\sum_{i = 0}^{n}\binom{n - i}{n - r} = \frac{1}{\binom{n}{r}}\sum_{j = 0}^{n}\binom{j}{n - r} = \frac{1}{\binom{n}{r}}\binom{n + 1}{n - r + 1} = \frac{n + 1}{n - r + 1},
		\end{align*}
		where the second last equality follows from the well-known Hockey Stick Identity $\sum_{k = i}^{n}\binom{k}{i} = \binom{n + 1}{i + 1}$  (see, e.g., \cite{jones1996generalized}).
	\end{IEEEproof}
	
	Now we are ready to prove Lemma \ref{lem:formula}.
	\begin{IEEEproof}[Proof of Lemma \ref{lem:formula}]
		We begin by analyzing the quantity $\gamma_i^s(G)$ for $i\in [k]$.
		Given $\cV\leq \bbF_q^k$, let 
		$$f(\cV) = \binom{nw_{\cV}(G)}{s}$$
		denote the number of $s$-element sets of columns of $G$ whose column vectors lie in $\cV$, and let
		$$g(\cV) = |\{\{\bh_1,\ldots,\bh_s\}:\bh_j\in\cV, \langle\bh_1,\ldots,\bh_s\rangle = \cV\}|$$
		denote the number of $s$-element sets of columns of $G$ whose column vectors lie in $\cV$ and span $\cV$.
		By partitioning subsets according to their span, we have
		$$f(\cV) = \sum_{\cU\leq\cV} g(\cU).$$
		
		Using the Möbius inversion formula for the lattice of subspaces of $\bbF_q^k$ (see, e.g., \cite[Propositions 3.7.1 and Example 3.10.2]{stanley2011enumerative}), let $r = \mathrm{dim}(\cV)$. Then we have
		\begin{align*}
			g(\cV) = \sum_{\cU\leq\cV}(-1)^{r - \mathrm{dim}(\cU)}q^{\binom{r - \mathrm{dim}(\cU)}{2}}f(\cU) = \sum_{t = 1}^{r}\sum_{\substack{\cU\leq \cV\\ \mathrm{dim}(\cU) = t}}(-1)^{r- t}q^{\binom{r- t}{2}}\binom{nw_{\cU}(G)}{s}.
		\end{align*}
		
		For any $s$-element set $S$ whose span does not contain $\be_i$, we have $\be_i\notin \langle S\rangle$. Counting such sets according to their $\mathbb{F}_q$-span yields that
		\begin{align*}
			\gamma_i^s(G) = &\sum_{\be_i\notin \cV}g(\cV)\\
			= &\sum_{r = 1}^{k - 1}\sum_{\be_i\notin\cV\in\bbS_{q, k}^r}\sum_{t = 1}^{r}\sum_{\cU\leq \cV, \mathrm{dim}(\cU) = t}(-1)^{r- t}q^{\binom{r-t}{2}}\binom{nw_{\cU}(G)}{s}\\
			= &\sum_{t = 1}^{k - 1}\sum_{\be_i\notin\cU\in\bbS_{q, k}^t}\sum_{r = t}^{k - 1}\sum_{\substack{\be_i\notin\cV, \cU\leq \cV,\\ \mathrm{dim}(\cV) = r}}(-1)^{r- t}q^{\binom{r- t}{2}}\binom{nw_{\cU}(G)}{s}\\
			= &\sum_{t = 1}^{k - 1}\sum_{\be_i\notin\cU\in\bbS_{q, k}^t}\sum_{r = t}^{k - 1}(-1)^{r- t}q^{\binom{r- t}{2}}\left(\qbin{k - t}{r - t} - \qbin{k - t - 1}{r - t - 1}\right)\binom{nw_{\cU}(G)}{s},
		\end{align*}
		where the last equality follows from the observation that the number of $r$-dimensional subspaces containing $\cU$ but not containing $\be_i$ is given by $\qbin{k - t}{r - t} - \qbin{k - t - 1}{r - t - 1}$.
		
		By Lemma \ref{lem:alpha}, the expectation $\bbE[\tau_i(G)]$ satisfies
		\begin{align*}
			\bbE[\tau_i(G)] 
			&= nH_n - \sum_{s = 1}^{n - 1}\frac{\alpha_i^s(G)}{\binom{n-1}{s}}\\
			&= nH_n - \sum_{s = 1}^{n - 1}\frac{\binom{n}{s}}{\binom{n-1}{s}} + \sum_{s = 1}^{n - 1}\frac{\gamma_i^s(G)}{\binom{n-1}{s}}\\
			&= nH_n - \sum_{s = 1}^{n - 1}\frac{n}{n-s} + \sum_{s = 1}^{n - 1}\frac{\gamma_i^s(G)}{\binom{n-1}{s}}\\
			&= 1 + \sum_{s = 1}^{n - 1}\frac{\gamma_i^s(G)}{\binom{n-1}{s}},
		\end{align*}       
		where the last equality follows from $$\sum_{s=1}^{n-1} \frac{n}{n-s} = nH_n - 1.$$
		Applying Claim~\ref{cla:combinatorial formula} and subtracting the $s=0$ term, we obtain
		\begin{align*}
			\sum_{s = 1}^{n - 1}\frac{\binom{nw_{\cU}(G)}{s}}{\binom{n - 1}{s}} = \frac{1}{1 - w_{\cU}(G)} - 1 = \frac{w_{\cU}(G)}{1 - w_{\cU}(G)}.
		\end{align*}
		Combining these results yields
		\begin{align*}
			\bbE[\tau_i(G)] = 1 + \sum_{t = 1}^{k - 1}\sum_{\substack{\be_i\notin\cU\\\cU\in\bbS_{q, k}^t}}\sum_{r = t}^{k - 1}(-1)^{r- t}q^{\binom{r- t}{2}}\left(\qbin{k - t}{r - t} - \qbin{k - t - 1}{r - t - 1}\right)\frac{w_{\cU}(G)}{1 - w_{\cU}(G)}.
		\end{align*}
	\end{IEEEproof}
	
	Building on Lemma \ref{lem:formula}, we can efficiently determine the maximum and average expected number of samples. For any subspace $\cU \leq \mathbb{F}_q^k$, let $h(\cU) = |\{i \in [k] : \be_i \notin \cU\}|$ denote the number of standard basis vectors not contained in $\cU$.
	
	\begin{corollary}\label{cor:average max}
		For any generator matrix $G\in\bbF_q^{k\times n}$, it holds that
		\begin{align*}
			T_{\max}(G) &= \max_{i\in [k]}\left\{1 + \sum_{t = 1}^{k - 1}\sum_{\substack{\be_i\notin\cU\\\cU\in\bbS_{q, k}^t}}\sum_{r = t}^{k - 1}(-1)^{r- t}q^{\binom{r- t}{2}}\left(\qbin{k - t}{r - t} - \qbin{k - t - 1}{r - t - 1}\right)\frac{w_{\cU}(G)}{1 - w_{\cU}(G)}\right\},\\
			T_{\mathrm{ave}}(G) &= 1 + \frac{1}{k}\sum_{t = 1}^{k-1}\sum_{\cU\in\bbS_{q, k}^t}h(\cU)\sum_{r = t}^{k - 1}(-1)^{r- t}q^{\binom{r- t}{2}}\left(\qbin{k - t}{r - t} - \qbin{k - t - 1}{r - t - 1}\right)\frac{w_{\cU}(G)}{1 - w_{\cU}(G)}.
		\end{align*}
	\end{corollary}

	\begin{IEEEproof}[Proof of Corollary \ref{cor:average max}]
		The expression for $T_{\max}(G)$ follows directly from Lemma~\ref{lem:formula}. We now derive the formula for $T_{\mathrm{ave}}(G)$. Summing the expectation $\bbE[\tau_i(G)]$ over all $i\in [k]$ yields
		\begin{align*}
			\sum_{i = 1}^{k}\bbE[\tau_i(G)] = &\sum_{i = 1}^{k}\left(1 + \sum_{t = 1}^{k - 1}\sum_{\be_i\notin\cU\in\bbS_{q, k}^t}\sum_{r = t}^{k - 1}(-1)^{r- t}q^{\binom{r- t}{2}}\left(\qbin{k - t}{r - t} - \qbin{k - t - 1}{r - t - 1}\right)\frac{w_{\cU}(G)}{1 - w_{\cU}(G)}\right)\\
			= &k + \sum_{t = 1}^{k - 1}\sum_{\cU\in\bbS_{q, k}^t}h(\cU)\sum_{r = t}^{k - 1}(-1)^{r- t}q^{\binom{r- t}{2}}\left(\qbin{k - t}{r - t} - \qbin{k - t - 1}{r - t - 1}\right)\frac{w_{\cU}(G)}{1 - w_{\cU}(G)},
		\end{align*}
		which implies that
		\begin{align*}
			T_{\mathrm{ave}}(G) = 1 + \frac{1}{k}\sum_{t = 1}^{k-1}\sum_{\cU\in\bbS_{q, k}^t}h(\cU)\sum_{r = t}^{k - 1}(-1)^{r- t}q^{\binom{r- t}{2}}\left(\qbin{k - t}{r - t} - \qbin{k - t - 1}{r - t - 1}\right)\frac{w_{\cU}(G)}{1 - w_{\cU}(G)}.
		\end{align*}
	\end{IEEEproof}
	 
	Algorithm~\ref{alg:e_i} provides a computational procedure for determining $\mathbb{E}[\tau_i(G)]$ for $i\in [k]$, based on Lemma~\ref{lem:formula}. Analogous algorithms for computing $T_{\mathrm{ave}}(G)$ and $T_{\max}(G)$ can be similarly derived from Corollary~\ref{cor:average max}. For fixed $q$ and $k$, Algorithm~\ref{alg:e_i} computes $\mathbb{E}[\tau_i(G)]$ in $O(n)$ time, which significantly improves upon the $O(n^2 2^n)$ complexity implied by a direct application of Lemma~\ref{lem:alpha}. The complexity analysis proceeds by noting that the algorithm must iterate through all subspaces of $\mathbb{F}_q^k$, the number of which is $O(q^{k^2})$. Since each subspace contains at most $q^k$ vectors, the overall time complexity of Algorithm~\ref{alg:e_i} is bounded by $O(n + q^{k^2 + 2k})$. This approach is efficient for small $q$ and $k$, but becomes computationally prohibitive for larger values due to the combinatorial explosion in the number of subspaces.

	\begin{example}
		Consider the case $q = 2$, $k = 3$. There are seven $1$-dimensional subspaces of $\mathbb{F}_2^3$ (omitting the zero vector for brevity):
		\begin{align*}
			\cU_1 &= \{\be_1\},~\cU_2 = \{\be_2\},~\cU_3 = \{\be_3\}, \cU_4 = \{\be_1 + \be_2\},~\cU_5 = \{\be_2 + \be_3\},~\cU_6 = \{\be_1 + \be_3\}, \cU_7 = \{\be_1 + \be_2 + \be_3\}.
		\end{align*}
		There are also seven $2$-dimensional subspaces of $\mathbb{F}_2^3$:
		\begin{align*}
			\cV_1 &= \{\be_1, \be_2, \be_1 + \be_2\},~\cV_2 = \{\be_2, \be_3, \be_2 + \be_3\},~\cV_3 = \{\be_1, \be_3, \be_1 + \be_3\},~\cV_4 = \{\be_1, \be_2 + \be_3, \be_1 + \be_2 + \be_3\},\\ \cV_5 &= \{\be_2, \be_1 + \be_3, \be_1 + \be_2 + \be_3\},~\cV_6 = \{\be_3, \be_1 + \be_2, \be_1 + \be_2 + \be_3\},~\cV_7 = \{\be_1 + \be_2, \be_2 + \be_3, \be_1 +\be_3\}.
		\end{align*}
		It follows that for $t = \dim(\cU)$, we have
		\[
		\sum_{r = 1}^{k - 1}(-1)^{r - t}q^{\binom{r- t}{2}}\left(\qbin{k - t}{r - t} - \qbin{k - t - 1}{r - t - 1}\right) = \begin{cases}
			-1, & \text{if } t = 1,\\
			1, & \text{if } t = 2.
		\end{cases}
		\]
		Therefore, for any generator matrix $G \in \mathbb{F}_2^{3 \times n}$, we have
		\begin{align*}
			T_{\mathrm{ave}}(G) = 1 + \frac{1}{3}\left[-2\sum_{i = 1}^3\frac{w_{\cU_i}(G)}{1 - w_{\cU_i}(G)} -3\sum_{i = 4}^7\frac{w_{\cU_i}(G)}{1 - w_{\cU_i}(G)} +\sum_{i = 1}^3\frac{w_{\cV_i}(G)}{1 - w_{\cV_i}(G)}+2\sum_{i = 4}^6\frac{w_{\cV_i}(G)}{1 - w_{\cV_i}(G)} + 3\frac{w_{\cV_7}(G)}{1 - w_{\cV_7}(G)}\right].
		\end{align*}
		Consider the Hamming code with generator matrix
		\[
			G = \begin{bmatrix}
				1 & 0 & 0 & 1 & 1 & 0 & 1\\
				0 & 1 & 0 & 1 & 0 & 1 & 1\\
				0 & 0 & 1 & 0 & 1 & 1 & 1
			\end{bmatrix}.
		\]
		For this matrix, every nonzero vector in $\mathbb{F}_2^3$ appears exactly once among the columns, so we have $w_{\cU_i}(G) = 1/7$ for all $i \in [7]$ and $w_{\cV_i}(G) = 3/7$ for all $i \in [7]$. Substituting these values yields
		\begin{align*}
			T_{\mathrm{ave}}(G) = 1 + \frac{1}{3}\left[-18\cdot \frac{1/7}{1 - 1/7} + 12\cdot \frac{3/7}{1 - 3/7}\right] = 3.
		\end{align*}
	\end{example}
	
	\begin{algorithm}
		\SetKwInOut{Input}{Input}\SetKwInOut{Output}{Output}
		\caption{Algorithm for $\bbE[\tau_i(G)]$}\label{alg:e_i}
		\Input{$G = [\bg_1, \ldots, \bg_n]\in\bbF_q^{k\times n}$, $i\in [k]$}
		\Output{$\bbE[\tau_i(G)]$}
		\BlankLine
		\For{$\bu\in\bbF_q^k$}{
			$w_{\bu} \gets 0$\;
		}
		\For{$j = 1$ \KwTo $n$}{
			$w_{\bg_j} \gets w_{\bg_j} + \frac{1}{n}$\;
		}
		$E \gets 1$\;
		\For{$t = 1$ \KwTo $k - 1$}{
			\For{$\be_i\notin \cU\in\bbS_{q, k}^{t}$}{
				$w_{\cU}(G) \gets 0$\;
				\For{$\bu\in\cU$}{
					$w_{\cU}(G) \gets w_{\cU}(G) + w_{\bu}$\;
				}
				\For{$r = t$ \KwTo $k - 1$}{
					$E \gets E + (-1)^{r- t}q^{\binom{r- t}{2}}\left(\qbin{k - t}{r - t} - \qbin{k - t - 1}{r - t - 1}\right)\frac{w_{\cU}(G)}{1 - w_{\cU}(G)}$\;
				}
			}
		}
		\Return{$E$}
	\end{algorithm}

	\section{Constructions of Generator Matrices}\label{sec:upper bounds}
	
	In this section, we examine constructions of generator matrices that achieve low average and maximum expected numbers of samples. With a slight abuse of notation, once we determine a desired weight distribution $(w_{\bu})_{\bu\in\bbF_q^k}$, we can construct $G$ by including $nw_{\bu}$ columns of $\bu$ when $nw_{\bu}$ is a positive integer. Lemma \ref{lem:brute-force search} provides explicit formulas for $T_{\max}(q, n, k)$ and $T_{\mathrm{ave}}(q, n, k)$ in terms of the weight distribution $(w_{\bu})_{\bu\in\bbF_q^k}$. 
	
	\begin{lemma}\label{lem:brute-force search}
		For any $q, n, k$ with $n\geq k$, let 
        \begin{align*}
             S = \Big\{&(w_{\bu})_{\bu\in\bbF_q^k}:~w_{\bu}\in\left\{\frac{j}{n}:0\leq j\leq n\right\}, \sum_{\bu\in\bbF_q^k}w_{\bu} = 1, \sum_{\bu\in\cU}w_{\bu} < 1,\forall~\cU\in\bbS_{q, k}^{k - 1}\Big\}
        \end{align*}
        be the set of all possible weight distributions of generator matrices in $\bbF_q^{k\times n}$. Then, for any $q, n, k$ with $n\geq k$, we have
        \begin{align*}
            T_{\max}(q , n , k)\ = \hspace{-5pt}\min_{(w_{\bu})_{\bu\in\bbF_q^k}\in S} \max_{i\in [k]}\Bigg\{1 + \sum_{t = 1}^{k - 1}\sum_{\substack{\be_i\notin\cU\\\cU\in\bbS_{q, k}^t}}
            \sum_{r = t}^{k - 1}(-1)^{r- t} \cdot q^{\binom{r- t}{2}}\left(\qbin{k - t}{r - t} - \qbin{k - t - 1}{r - t - 1}\right)\frac{\sum_{\bu\in\cU}w_{\bu}}{1 - \sum_{\bu\in\cU}w_{\bu}}\Bigg\},
        \end{align*}
        \begin{align*}
            T_{\mathrm{ave}}(q, n, k) = \min_{(w_{\bu})_{\bu\in\bbF_q^k}\in S}\Bigg\{1+ \sum_{t = 1}^{k-1}\sum_{\cU\in\bbS_{q, k}^t}\frac{h(\cU)}{k}\sum_{r = t}^{k - 1}(-1)^{r- t} \cdot q^{\binom{r- t}{2}}\left(\qbin{k - t}{r - t} - \qbin{k - t - 1}{r - t - 1}\right)\frac{\sum_{\bu\in\cU}w_{\bu}}{1 - \sum_{\bu\in\cU}w_{\bu}}\Bigg\}.
        \end{align*}
	\end{lemma}
	\begin{IEEEproof}[Proof of Lemma \ref{lem:brute-force search}]
		By Lemma \ref{lem:formula}, for any generator matrix $G$, it holds that $T_{\max}(G)$ is given by
		\[
		T_{\max}(G) = \max_{i\in [k]}\left\{1 + \sum_{t = 1}^{k - 1}\sum_{\be_i\notin\cU\in\bbS_{q, k}^t}\sum_{r = t}^{k - 1}(-1)^{r- t}q^{\binom{r- t}{2}}\left(\qbin{k - t}{r - t} - \qbin{k - t - 1}{r - t - 1}\right)\frac{w_{\cU}(G)}{1 - w_{\cU}(G)}\right\}
		\]
		and $T_{\mathrm{ave}}(G)$ is expressed as
		\[
		T_{\mathrm{ave}}(G) = 1 + \frac{1}{k}\sum_{t = 1}^{k-1}\sum_{\cU\in\bbS_{q, k}^t}h(\cU)\sum_{r = t}^{k - 1}(-1)^{r- t}q^{\binom{r- t}{2}}\left(\qbin{k - t}{r - t} - \qbin{k - t - 1}{r - t - 1}\right)\frac{w_{\cU}(G)}{1 - w_{\cU}(G)}.
		\]
	Since both quantities depend solely on the choice of the generator matrix $G$, and the set of such matrices is finite for fixed $q$, $n$, and $k$, it suffices to search exhaustively over all possible generator matrices $G \in \mathbb{F}_q^{k \times n}$.

    For a matrix $G = [\bg_1, \ldots, \bg_n] \in \mathbb{F}_q^{k \times n}$, the rank of $G$ is $k$ if and only if, for every $(k-1)$-dimensional subspace $\cU \in \bbS_{q, k}^{k-1}$, there exists at least one column $\bg_i \notin \cU$. Equivalently, this condition implies that $w_{\cU}(G) < 1$ for all $\cU \in \bbS_{q, k}^{k-1}$. Since $G$ has $n$ columns, each proportion $w_{\bu}(G)$ takes values in the set $\{\frac{j}{n} : 0 \leq j \leq n\}$. This completes the proof.
\end{IEEEproof}
	
	Lemma \ref{lem:brute-force search} reduces the problem of optimizing over generator matrices to an optimization over weight distributions, thereby enabling exhaustive search for small parameters. Table \ref{tab:q=2k=3} presents the values of $T_{\max}(q, n, k)$ and $T_{\mathrm{ave}}(q, n, k)$ along with their corresponding weight distributions $(w_{\bu})_{\bu\in\mathbb{F}_2^3}$ for $q=2$, $k=3$, and $4 \leq n \leq 8$.
	
	{\begin{table*}[htbp]
        \setlength{\tabcolsep}{5pt}
		\centering
		\caption{The Values of $T_{\max}(q = 2, n, k = 3)$ and $T_{\mathrm{ave}}(q = 2, n, k = 3)$ for $4\leq n\leq 8$.}\label{tab:q=2k=3}
		\renewcommand{\arraystretch}{1.2}
		\begin{tabular}{|c|cc|cc|cc|cc|cc|}
			\hline
			\multirow{2}{*}{$w_{\bu}$} &
			\multicolumn{2}{c|}{$n=4$} &
			\multicolumn{2}{c|}{$n=5$} &
			\multicolumn{2}{c|}{$n=6$} &
			\multicolumn{2}{c|}{$n=7$} &
			\multicolumn{2}{c|}{$n=8$} \\ \cline{2-11}
			& $T_{\max}$ & $T_{\mathrm{ave}}$ & $T_{\max}$ & $T_{\mathrm{ave}}$ & $T_{\max}$ & $T_{\mathrm{ave}}$ & $T_{\max}$ & $T_{\mathrm{ave}}$ & $T_{\max}$ & $T_{\mathrm{ave}}$ \\ \hline
			Value &
			$3$ & $3$ &
			$\tfrac{37}{12} \approx 3.083$ & $\tfrac{53}{18} \approx 2.944$ &
			$3$ & $\tfrac{89}{30} \approx 2.967$ &
			$\tfrac{43}{15} \approx 2.867$ & $\tfrac{43}{15} \approx 2.867$ &
			$\tfrac{313}{105} \approx 2.981$ & $\tfrac{299}{105} \approx 2.848$ \\ \hline
			$w_{\be_1}$ &
			$1/4$ & $1/4$ &
			$1/5$ & $1/5$ &
			$1/6$ & $1/6$ &
			$2/7$ & $1/7$ &
			$1/4$ & $1/4$ \\
			$w_{\be_2}$ &
			$1/4$ & $1/4$ &
			$1/5$ & $1/5$ &
			$1/6$ & $1/6$ &
			$2/7$ & $2/7$ &
			$1/4$ & $1/4$ \\
			$w_{\be_3}$ &
			$1/4$ & $1/4$ &
			$1/5$ & $1/5$ &
			$1/6$ & $1/3$ &
			$2/7$ & $2/7$ &
			$1/4$ & $1/4$ \\
			$w_{\be_{1} + \be_2}$ &
			$0$ & $0$ &
			$0$ & $0$ &
			$0$ & $1/6$ &
			$0$ & $1/7$ &
			$0$ & $0$ \\
			$w_{\be_{2}+\be_3}$ &
			$0$ & $0$ &
			$0$ & $1/5$ &
			$1/6$ & $0$ &
			$0$ & $0$ &
			$1/8$ & $1/8$ \\
			$w_{\be_{1} + \be_3}$ &
			$0$ & $0$ &
			$1/5$ & $1/5$ &
			$1/6$ & $1/6$ &
			$0$ & $1/7$ &
			$1/8$ & $1/8$ \\
			$w_{\be_{1}+\be_2+\be_3}$ &
			$1/4$ & $1/4$ &
			$1/5$ & $0$ &
			$1/6$ & $0$ &
			$1/7$ & $0$ &
			$0$ & $0$ \\ \hline
		\end{tabular}
	\end{table*}}
	
	Several observations follow from Table~\ref{tab:q=2k=3}. First, neither $T_{\max}(q = 2, n, k = 3)$ nor $T_{\mathrm{ave}}(q = 2, n, k = 3)$ decreases monotonically with $n$, and the matrices minimizing the average expected number of samples may differ from those minimizing the maximum expected number of samples. It has been shown that $\gamma T_{\max}(q, n, k) \leq T_{\max}(q, n, \gamma k)$ and $\gamma T_{\mathrm{ave}}(q, n, k) \leq T_{\mathrm{ave}}(q, n, \gamma k)$ holds for any integer $\gamma \geq 1$ \cite{daniella2023cover}. However, to the best of our knowledge, it remains an open problem whether there exists an integer $N = N(q, k)$ such that for all $n > N$, both $T_{\max}(q, n, k)$ and $T_{\mathrm{ave}}(q, n, k)$ are non-increasing in $n$. Corollary \ref{cor:monotone} follows directly from Lemma \ref{lem:brute-force search} and demonstrates that monotonicity holds under a divisibility condition.
	\begin{corollary}\label{cor:monotone}
		For $n$, $n'$ with $n' \mid n$, we have $T_{\mathrm{ave}}(q, n, k) \leq T_{\mathrm{ave}}(q, n', k)$ and $T_{\max}(q, n, k) \leq T_{\max}(q, n', k)$.
	\end{corollary}
	
	When $n$ is large, finding the optimal construction via exhaustive search becomes computationally infeasible. Lemma \ref{lem:construction} provides a useful tool for analyzing upper bounds on $T_{\mathrm{ave}}(q, k)$ and $T_{\max}(q, k)$.
	\begin{lemma}\label{lem:construction}
		For any rational numbers $w_1, \ldots, w_k$ satisfying $0 \leq w_{i} \leq 1$ and $\sum_{i = 1}^{k}(q-1)^{i}\binom{k}{i}w_{i} = 1$, let $G$ be a matrix such that $w_{\bu}(G) = w_{\mathrm{wt}(\bu)}$ for all $\bu \in \bbF_q^k$, then
		\begin{align*}
			T_{\mathrm{ave}}(q, k) \leq T_{\max}(q, k) \leq 1 + \frac{1}{k}\sum_{t = 1}^{k-1}\sum_{\cU\in\bbS_{q, k}^t}h(\cU)\sum_{r = t}^{k - 1}(-1)^{r- t}q^{\binom{r- t}{2}}\left(\qbin{k - t}{r - t} - \qbin{k - t - 1}{r - t - 1}\right)\frac{w_{\cU}(G)}{1 - w_{\cU}(G)}.
		\end{align*}
	\end{lemma}
	\begin{IEEEproof}[Proof of Lemma \ref{lem:construction}]
		Choose rational $w_1, \ldots, w_k$ such that $0 \leq w_{i} \leq 1$ and satisfy the condition $\sum_{i = 1}^{k}(q-1)^{i}\binom{k}{i}w_{i} = 1$. Construct a generator matrix $G$ where the number of columns corresponding to each $\bu \in \bbF_q^k$ is given by $w_{\mathrm{wt}(\bu)}n$, ensuring that $w_{\bu}(G) = w_{\mathrm{wt}(\bu)}$. By applying Lemma~\ref{lem:formula} and exploiting the symmetry among the coordinates, we have:
		\begin{align*}
			\bbE[\tau_1(G)] = \ldots = \bbE[\tau_k(G)] = 1 + \frac{1}{k}\sum_{t = 1}^{k-1}\sum_{\cU\in\bbS_{q, k}^t}h(\cU)\sum_{r = t}^{k - 1}(-1)^{r- t}q^{\binom{r- t}{2}}\left(\qbin{k - t}{r - t} - \qbin{k - t - 1}{r - t - 1}\right)\frac{w_{\cU}(G)}{1 - w_{\cU}(G)},
		\end{align*}
		which implies that 
		\begin{align*}
			T_{\mathrm{ave}}(q, k) \leq T_{\max}(q, k) \leq 1 + \frac{1}{k}\sum_{t = 1}^{k-1}\sum_{\cU\in\bbS_{q, k}^t}h(\cU)\sum_{r = t}^{k - 1}(-1)^{r- t}q^{\binom{r- t}{2}}\left(\qbin{k - t}{r - t} - \qbin{k - t - 1}{r - t - 1}\right)\frac{w_{\cU}(G)}{1 - w_{\cU}(G)}.
		\end{align*}
	\end{IEEEproof}
	
	\begin{example}
		Consider the case $q = 2$ and $k = 3$. Suppose rational numbers $0 \leq w_1, w_2, w_3 \leq 1$ satisfy $\sum_{i = 1}^{3}\binom{3}{i}w_{i} = 1$ and let $G$ be a matrix such that $w_{\bu}(G) = w_{\mathrm{wt}(\bu)}$. By Lemma \ref{lem:construction}, we have
		
		\begin{align*}
			T_{\max}(q = 2, k = 3)\leq -\frac{2}{1 - w_1} - \frac{3}{1 - w_2} - \frac{1}{1 - w_3}+ \frac{1}{1 - 2w_1 - w_2} + \frac{2}{1 - w_1 - w_2 - w_3}+ \frac{1}{1 - 3w_2} + 3.
		\end{align*}
		By numerical optimization, we have $T_{\mathrm{ave}}(q = 2, k = 3) \leq T_{\max}(q = 2, k = 3) \leq 0.9263 \cdot 3$. This result improves the upper bound $\frac{8\sqrt{3}\pi - 18}{27} k \approx 0.9456 k$ given in \cite{bodur2025random} for $q=2$.
	\end{example}
	Next, we consider the regime where $q$ is large. Lemma~\ref{lem:weight distribution 3} provides an upper bound that is valid for all $q$.
	
	\begin{lemma}\label{lem:weight distribution 3}
		For any rational numbers $0 \leq w_1, w_2, w_3 \leq 1$ such that $3(q - 1)w_1 + 3(q - 1)^2w_2 + (q - 1)^3w_3 = 1$, it holds that
		\begin{align*}
			T_{\max}(q, k = 3)\leq &-\frac{2(q - 1)}{1 - (q - 1)w_1} - \frac{3(q - 1)^2}{1 - (q - 1)w_2} - \frac{(q - 1)^3}{1 - (q - 1)w_3} +\frac{2(q - 1)}{1 - (q - 1)w_1 - (q - 1)w_2 - (q - 1)^2w_3} \\
			&+ \frac{(q - 1)^2}{1 - 3(q - 1)w_2 - (q - 1)(q - 2)w_3} + \frac{1}{1 - 2(q - 1)w_1 - (q - 1)^2w_2}+ (q - 1)^3 + 2(q - 1)^2.
		\end{align*}
	\end{lemma}
	Before proving Lemma \ref{lem:weight distribution 3}, we first define the weight distribution of a subspace $\cU$. For a subspace $\cU\leq \bbF_q^k$, its weight distribution is defined as $\mathrm{wd}(\cU) = (a_1(\cU), \ldots, a_k(\cU))$, where $a_i(\cU) = |\{\bu\in \cU: \mathrm{wt}(\bu) = i\}|$ for $i\in [k]$.
	\begin{IEEEproof}[Proof of Lemma \ref{lem:weight distribution 3}]
		To prove this conclusion, we first enumerate the weight distributions and count the corresponding subspaces of $\mathbb{F}_q^3$. Table~\ref{tab:weight distribution 3} provides the number of subspaces of $\mathbb{F}_q^3$ for each weight distribution and dimension.
		\begin{table}[H]
			\centering
			\caption{Number of subspaces of $\bbF_q^3$ for each weight distribution and dimension}\label{tab:weight distribution 3}
			\begin{tabular}{|c|c|c|}
				\hline
				Dimension & Weight Distribution & Number of Subspaces \\\hline
				1 & $(q - 1, 0, 0)$ & $3$ \\\hline
				1 & $(0, q - 1, 0)$ & $3(q - 1)$ \\\hline
				1 & $(0, 0, q - 1)$ & $(q - 1)^{2}$ \\\hline
				2 & $(2(q - 1), (q - 1)^2, 0)$ & $3$ \\\hline
				2 & $(q - 1, q - 1, (q - 1)^2)$ & $3(q - 1)$ \\\hline
				2 & $(0, 3(q - 1), (q - 1)(q - 2))$ & $(q - 1)^{2}$ \\\hline
			\end{tabular}
		\end{table}

		We begin by analyzing the $1$-dimensional subspaces of $\bbF_q^3$. For a space $\cU\in\bbS_{q,3}^1$, the weight distribution is $\mathrm{wd}(\cU) = (q - 1, 0, 0)$ if $\cU = \langle \be_i\rangle$ for $i\in [3]$. If $\cU = \langle \be_i + a \be_j\rangle$ for $i\neq j\in [3]$ and $a\neq 0$, then $\mathrm{wd}(\cU) = (0, q - 1, 0)$. All remaining one-dimensional subspaces yield a weight distribution of $(0,0,q-1)$.
		
		Next, we analyze the $2$-dimensional spaces. There are three $2$-dimensional subspaces in $\bbS_{q,3}^2$ that contain two standard basis vectors. For a space $\cU\in \bbS_{q,3}^2$ given by $\cU = \langle \be_i, \be_j\rangle$ for $i\neq j\in [3]$, we have $\mathrm{wd}(\cU) = (2(q - 1), (q - 1)^2, 0)$.
		
		For each standard basis vector $\be_i$, there are $\qbin{2}{1} - 2 = q - 1$ $2$-dimensional subspaces containing $\be_i$ without a second standard basis vector, yielding a total of $3(q - 1)$ such subspaces. If $\cU$ contains exactly one standard basis vector, say $\be_1$, then $\cU = \langle\be_1, a\be_2 + b\be_3\rangle$ where $a, b\neq 0$, giving $\mathrm{wd}(\cU) = (q - 1, q - 1, (q - 1)^2)$.
		
		The remaining $\qbin{3}{2} - 3 - 3(q - 1) = (q - 1)^2$ $2$-dimensional subspaces contain no standard basis vectors. For such a subspace $\cU$, we have $\cU = \langle \be_1 + a\be_2, \be_2 + b\be_3\rangle$ for $a, b\neq 0$. The number of vectors of weight $2$ is $3(q - 1)$, while the number of vectors of weight $3$ is $(q - 1)^2 - 3(q - 1) = (q - 1)(q - 2)$ in total.
		
		A direct computation shows that 
		\begin{align*}
			\sum_{r = t}^{k - 1}(-1)^{r- t}q^{\binom{r- t}{2}}\left(\qbin{k - t}{r - t} - \qbin{k - t - 1}{r - t - 1}\right) = \begin{cases}
				1, & t = 2,\\
				1 - q, & t = 1.
			\end{cases}
		\end{align*}
		For any $0\leq w_1, w_2, w_3\leq 1$ satisfying $3(q - 1)w_1 + 3(q - 1)^2w_2+(q - 1)^3w_3 = 1$, let $G$ be a matrix such that $w_{\bu}(G) = w_{\mathrm{wt}(\bu)}$. By Lemma \ref{lem:formula}, we obtain
		\begin{align*}
			T_{\mathrm{ave}}(G) = &1 + \frac{1}{3}\sum_{t = 1}^{2}\sum_{\cU\in\bbS_{q, k}^t}h(\cU)\sum_{r = t}^{2}(-1)^{r- t}q^{\binom{r- t}{2}}\left(\qbin{k - t}{r - t} - \qbin{k - t - 1}{r - t - 1}\right)\frac{w_{\cU}(G)}{1 - w_{\cU}(G)}\\
            =&-\frac{2(q - 1)}{1 - (q - 1)w_1} - \frac{3(q - 1)^2}{1 - (q - 1)w_2} - \frac{(q - 1)^3}{1 - (q - 1)w_3} +\frac{2(q - 1)}{1 - (q - 1)w_1 - (q - 1)w_2 - (q - 1)^2w_3} \\
			&+ \frac{(q - 1)^2}{1 - 3(q - 1)w_2 - (q - 1)(q - 2)w_3} + \frac{1}{1 - 2(q - 1)w_1 - (q - 1)^2w_2}+ (q - 1)^3 + 2(q - 1)^2.
		\end{align*}
        The fact that $T_{\mathrm{ave}}(G) = T_{\max}(G)$ completes the proof.
	\end{IEEEproof}
	For small values of $q$, it is computationally feasible to search for $w_1, w_2,$ and $w_3$ to obtain an upper bound for $T_{\max}(q, k = 3)$ Table \ref{tab:search_results} lists some upper bounds of $T_{\max}(q, k=3)$ and corresponding weights $w_i$ for field sizes $2\leq q\leq 8$ by computer search. Next, we investigate the asymptotic upper bounds on $\limsup_{q\to\infty}T_{\max}(q, 3)$.

	\begin{table}[htbp]
		\centering
		\caption{The upper bounds of $T_{\max}(q, k=3)$ and the corresponding weights $w_i$ for field sizes $2\leq q\leq 8$}\label{tab:search_results}
		\begin{tabular}{|l|c|c|c|c|c|c|}
		\hline
		$q$ & $2$ & $3$ & $4$ & $5$ & $7$ & $8$ \\ \hline
		$T_{\max}$ & 2.7789 & 2.7240 & 2.7006 & 2.6878 & 2.6742 & 2.6702 \\ \hline
		$w_1$      & 0.2382 & 0.1057 & 0.0665 & 0.0482 & 0.0308 & 0.0261 \\ \hline
		$w_2$      & 0.0785 & 0.0232 & 0.0110 & 0.0065 & 0.0031 & 0.0023 \\ \hline
		$w_3$      & 0.0500 & 0.0109 & 0.0039 & 0.0017 & 0.0005 & 0.0003 \\ \hline
		\end{tabular}
	\end{table}
	
	\begin{theorem}\label{thm:k=3}
		When $k = 3$, it holds that 
		\[
		\limsup_{q\to\infty}T_{\max}(q, k = 3)\leq 0.8811 \cdot 3.
		\]
	\end{theorem}
	\begin{IEEEproof}[Proof of Theorem \ref{thm:k=3}]
		For convenience, let $x = (q - 1)w_1$, $y = (q - 1)w_2$, and $z = (q - 1)w_3$, where $0\leq x, y, z\leq 1$ and $3x + 3(q - 1)y+(q - 1)^2z = 1$. Let
		\begin{align*}
			F(x, y, z) = &-\frac{2(q - 1)}{1 - x} - \frac{3(q - 1)^2}{1 - y} - \frac{(q - 1)^3}{1 - z} + \frac{1}{1 - 2x - (q - 1)y} + (q - 1)^3\\
            & + \frac{2(q - 1)}{1 - x - y - (q - 1)z}+ \frac{(q - 1)^2}{1 - 3y - (q - 2)z} + 2(q - 1)^2\\
			= & (q - 1)^3\left(1 - \frac{1}{1-z}\right) + \frac{1}{1-2x-(q-1)y} + (q - 1)^2\left(2 + \frac{1}{1-3y-(q-2)z} - \frac{3}{1-y}\right)\\
			& + (q - 1)\left(\frac{2}{1-x-y-(q - 1)z} - \frac{2}{1-x}\right).
		\end{align*}
		Let $x = \frac{1}{3} - \frac{\lambda}{3} - \mu$, $y = \frac{\mu}{q - 1}$, and $z = \frac{\lambda}{(q - 1)^2}$, where $\lambda$ and $\mu$ are constants to be determined. By Taylor expansion, we have
		\begin{align*}
			F\left(\frac{1}{3} - \frac{\lambda}{3} - \mu, \frac{\mu}{q - 1}, \frac{\lambda}{(q - 1)^2}\right) = &(q - 1)^3\left(1 - \frac{1}{1-\frac{\lambda}{(q - 1)^2}}\right) + \frac{1}{\frac{1}{3} + \frac{2\lambda}{3} + \mu}\\
			&+ (q - 1)^2\left(2 + \frac{1}{1-3\frac{\mu}{q - 1}-(q-2)\frac{\lambda}{(q - 1)^2}} - \frac{3}{1-\frac{\mu}{q - 1}}\right)\\
			& + (q - 1)\left(\frac{2}{\frac{2}{3} + \frac{\lambda}{3} + \mu-\frac{\lambda + \mu}{q - 1}} - \frac{2}{\frac{2}{3} + \frac{\lambda}{3} + \mu}\right)\\
			= &(q - 1)^3\left(\frac{-\lambda}{(q - 1)^2}\right) + \frac{1}{\frac{1}{3} + \frac{2\lambda}{3} + \mu}\\
			&+ (q - 1)^2\left(\hspace{-0.1ex}\frac{(q-2)\lambda}{(q - 1)^2}\hspace{-0.1ex} + \hspace{-0.1ex}\frac{6\mu^2}{(q - 1)^2}\hspace{-0.1ex}+\hspace{-0.1ex}\frac{6(q - 2)\lambda\mu}{(q - 1)^3}\hspace{-0.1ex}+\hspace{-0.1ex}\frac{(q-2)^2\lambda^2}{(q - 1)^4}\hspace{-0.3ex}\right)\\
			& + (q - 1)\left(\frac{2\frac{\lambda + \mu}{q - 1}}{\left(\frac{2}{3} + \frac{\lambda}{3} + \mu\right)\left(\frac{2}{3} + \frac{\lambda}{3} + \mu-\frac{\lambda + \mu}{q - 1}\right)}\right) +  O\left(\frac{1}{q}\right)\\
			= &-\lambda + 6\mu^2+6\lambda\mu+\lambda^2+\frac{2(\lambda+\mu)}{\left(\frac{2}{3} + \frac{\lambda}{3} + \mu\right)^2} + \frac{1}{\frac{1}{3} + \frac{2\lambda}{3} + \mu}~(q\to\infty).
	\end{align*}
	By a numerical computer-assisted search, it can be found that when $\lambda \approx 0.06679$ and $\mu \approx 0.1509$, we have that $$F\left(\frac{1}{3} - \frac{\lambda}{3} - \mu, \frac{\mu}{q - 1}, \frac{\lambda}{(q - 1)^2}\right) < 2.6433 = 0.8811\cdot 3,$$ which completes the proof.
	\end{IEEEproof}
	Theorem \ref{thm:k=3} improves the upper bounds for $k=3$ in \cite{gruica2024combinatorial,gruica2024geometry,bodur2025random}.
	To the best of our knowledge, this is the state-of-the-art upper bound for $k = 3$. With a similar method, we can also derive an upper bound for $k=4$. The following theorem summarizes the result, which improves the upper bound for $k=4$ in \cite{boruchovsky2025making}, which is approximately $0.86375 \cdot 4$.
	\begin{theorem}\label{thm:k=4}
		When $k = 4$, it holds that 
		\[
		\limsup_{q\to\infty}T_{\max}(q, k = 4) < 0.862882 \cdot 4.
		\]
	\end{theorem}
	The proof of Theorem \ref{thm:k=4} is provided in Appendix \ref{app:k=4}. It is worth noting that the upper bounds in Theorems \ref{thm:k=3} and Theorem \ref{thm:k=4} are obtained by determining the weight distribution of all subspaces of $\bbF_q^k$. To the best of our knowledge, for larger values of $k$, it is still an open problem to determine the weight distribution of all subspaces of $\bbF_q^k$. We leave this as an interesting direction for future research.
	
	\section{Lower Bounds for the Expected Number of Samples}\label{sec:lower bounds}
	
	In this section, we study lower bounds for $T_{\max}(q, n, k)$. Prior work in~\cite{daniella2023cover} established that $T_{\max}(q, n, k)\geq \frac{k + 1}{2}$ and $T_{\max}(q, n, k)\geq n - \frac{n(n - k)}{k}(H_n - H_{n - k})$, with the latter bound being tight when $n = k$. In the remainder of this section, we assume $n\geq k + 1$. For $G\in\bbF_q^{k \times n}$ and $s\leq n - 1$, let $\zeta^s(G)$ denote the number of $s$-element column sets of $G$ whose $\mathbb{F}_q$-span is not a $\min\{s, k\}$-dimensional standard space. Intuitively, $\zeta^s(G)$ quantifies the number of column sets whose span is insufficient for recovering $\min\{s, k\}$ information strands.

	\begin{example}
		Let $q = 2, k = 2$, and $n = 5$. Consider the matrix in Example \ref{exa:1}
		\[
		G = [\bg_1, \bg_2, \bg_3, \bg_4, \bg_5] = \begin{bmatrix}
			1 & 0 & 1 & 0 & 1\\
			0 & 1 & 0 & 1 & 1
		\end{bmatrix}\in\bbF_2^{2\times 5}.
		\]
		Then $\zeta^1(G) = 1$, $\zeta^2(G) = 2$, and $\zeta^3(G) = \zeta^4(G) = 0$.
	\end{example}
	
	Lemma \ref{lem:lower bound zeta} shows a relationship between $\alpha_i^s(G)$ and $\zeta^s(G)$.
	
	\begin{lemma}\label{lem:lower bound zeta}
		For any generator matrix $G\in\bbF_q^{k \times n}$, we have 
		\[
		\sum_{i = 1}^{k}\alpha_i^s(G)\leq \min\{s, k\}\binom{n}{s} - \zeta^s(G).
		\]    
	\end{lemma}
	\begin{IEEEproof}[Proof of Lemma \ref{lem:lower bound zeta}]
		For any $s$-element set $S$ of columns of $G$, since $\mathrm{dim}(\langle S\rangle) \leq \min\{s, k\}$, it follows that $S$ can serve as a recovery set for at most $\min\{s, k\}$ standard basis vectors. Furthermore, if $\langle S\rangle$ is not a $\min\{s, k\}$-dimensional standard subspace, then $S$ can be a recovery set for at most $\min\{s, k\} - 1$ standard basis vectors. In total, we have
		\[
		\sum_{i = 1}^{k}\alpha_i^s(G) \leq \min\{s, k\}\binom{n}{s} - \zeta^s(G).
		\]
	\end{IEEEproof}
	Based on Lemma \ref{lem:lower bound zeta}, we can derive a lower bound on $\zeta^s(G)$ in  Lemma~\ref{lem:count}.
	\begin{lemma}\label{lem:count}
		Given a rank-$k$ generator matrix $G\in\bbF_{q}^{k\times n}$, let $r$ be the smallest positive integer such that some $r$-dimensional subspace contains $r + 1$ columns of $G$. For $r \geq 1$ and $r + 1 \leq s \leq k$, it holds that
		\[
		\zeta^s(G) \geq \binom{n-(r + 1)}{s - (r + 1)} + \sum_{t = 1}^{r - 1}\binom{r}{t - 1}\binom{k - r}{s - t}.
		\]
	\end{lemma}
	Claim~\ref{cla:tool} plays a crucial role in the proof of Lemma~\ref{lem:count}.
	\begin{claim}\label{cla:tool} 
		Given $\ell\geq 1$, for any $(r + \ell)$-dimensional subspace $\cW \leq \bbF_q^k$ and any $\ell$-dimensional subspace $\cV \leq \cW$, the number of $(t + \ell)$-dimensional standard subspaces of $\cW$ that contain $\cV$ is at most $\binom{r}{t}$.
	\end{claim}

	\begin{IEEEproof}[Proof of Claim \ref{cla:tool}]
		Since $\dim(\cV) = \ell$, it follows that $|\mathrm{supp}(\cV)| \ge \ell$. For a $(t + \ell)$-dimensional standard subspace $\cV'$ to satisfy the condition $\cV \subseteq \cV' \subseteq \cW$, it holds that $\mathrm{supp}(\cV) \subseteq \mathrm{supp}(\cV')$ and for every $e_i \in \mathrm{supp}(\cV')$, the standard basis vector $e_i$ must belong to $\cW$.

		Let $I_{\cW} = \{i \in [k] : e_i \in \cW\}$ be the set of indices of standard basis vectors contained in $\cW$. Note that $|I_{\cW}| \le \dim(\cW) = r+\ell$ and $\mathrm{supp}(\cV) \subseteq \mathrm{supp}(\cV') \subseteq I_{\cW}$. Consequently, determining a valid $\cV'$ is equivalent to choosing the remaining $(t+\ell) - |\mathrm{supp}(\cV)|$ elements of $\mathrm{supp}(\cV')$ from the set $I_{\cW} \setminus \mathrm{supp}(\cV)$. 

		Given that $|\mathrm{supp}(\cV)| \ge \ell$ and $|I_{\cW}| \le r + \ell$, the number of such standard subspaces $\cV'$ is at most:
		$$\binom{|I_{\cW}| - |\mathrm{supp}(\cV)|}{(t+\ell) - |\mathrm{supp}(\cV)|} \le \binom{(r+\ell) - |\mathrm{supp}(\cV)|}{(t+\ell) - |\mathrm{supp}(\cV)|} \le \binom{(r+\ell) - \ell}{(t+\ell) - \ell} = \binom{r}{t}.$$
	\end{IEEEproof}
	Now we are ready to prove Lemma \ref{lem:count}.
	\begin{IEEEproof}[Proof of Lemma \ref{lem:count}]
		Let $G\in\bbF_q^{k\times n}$ be a rank-$k$ generator matrix, and let $r$ be the smallest positive integer such that some $r$-dimensional space contains $r + 1$ columns of $G$. Without loss of generality, assume $\cU = \langle\bg_1, \ldots, \bg_{r + 1}\rangle\in\bbS_{q, k}^r$. The following observations hold:
		\begin{itemize}
			\item By minimality of $r$, any $r$ columns from $\bg_1, \ldots, \bg_{r + 1}$ span $\cU$.
			\item Since $G$ has rank $k$, there exist $k - r$ columns among the remaining columns such that, together with $\cU$, they span $\bbF_q^k$.
		\end{itemize}
		
		With the observations above, we can derive lower bounds on $\zeta^s(G)$. We begin with the case $r \geq 2$. Without loss of generality, assume that $\cU = \langle\bg_1, \ldots, \bg_{r + 1}\rangle\in\bbS_{q, k}^r$ and $\langle\cU, \bg_{r + 2}, \ldots, \bg_{k + 1}\rangle = \bbF_q^k$. Consider any $\ell \leq k - r$ and $t\leq r - 1$. For a set of $\ell$ columns $V = \{\bg_{j_1}, \ldots, \bg_{j_\ell}\} \subset \{\bg_{r + 2}, \ldots, \bg_{k + 1}\}$, we assert that the number of $t$-element subsets $U \subset \{\bg_1, \ldots, \bg_{r + 1}\}$ for which $\langle U, V\rangle$ is a standard subspace is at most $\binom{r}{t}$.
		
		Suppose for contradiction that there exist $\binom{r}{t} + 1$ distinct $t$-element sets $U_1, \ldots, U_{\binom{r}{t} + 1} \subset \{\bg_1, \ldots, \bg_{r + 1}\}$ and $V = \{\bg_{r + 2}, \ldots, \bg_{r + \ell + 1}\} \subset \{\bg_{r + 2}, \ldots, \bg_{k + 1}\}$ such that $\langle U_i, V\rangle$ are all standard subspaces. By the minimality of $r$, we have $\mathrm{dim}(\langle U_i \cup U_j\rangle) > t$ for $i \neq j$, which implies that $\langle U_i, V\rangle \neq \langle U_j, V\rangle$ for $i \neq j$. By Claim \ref{cla:tool}, there are at most $\binom{r}{t}$ standard subspaces of $\cW$ that contain $\langle V\rangle$, which is a contradiction.
		
		Now observe that for any $s$-element set $S$ with $s \geq r + 1$ containing $U$, we have $\mathrm{dim}(\langle S\rangle) \leq s - 1$. Therefore, $S$ cannot span an $s$-dimensional standard subspace. In total, there are at least 
		\[
		\binom{n-(r + 1)}{s - (r + 1)} + \sum_{t = 1}^{r - 1}\binom{r}{t - 1}\binom{k - r}{s - t}
		\]
		sets of size $s$ that do not generate an $s$-dimensional standard subspace.

		When $r = 1$, it holds that $\sum_{t = 1}^{r - 1}\binom{r}{t - 1}\binom{k - r}{s - t} = 0$, which implies that the formula also holds for $r = 1$.
	\end{IEEEproof}
	
	Now we can provide the lower bound for $T_{\mathrm{ave}}(q, n, k)$. For simplicity, let
	$$S(r) = \sum_{s = 1}^{r - 1}\frac{\binom{n}{s} - \binom{k}{s}}{k\binom{n - 1}{s}} + \sum_{s = r}^k\frac{\binom{n - (r + 1)}{s - (r + 1)} + \sum_{t = 1}^{r - 1}\binom{r}{t - 1}\binom{k - r}{s - t}}{k\binom{n - 1}{s}}.$$
	
	\begin{theorem}\label{thm:lower bound k + 1}
		For any $q, n, k$ with $n\geq k + 1$, it holds that
		\begin{align*}
			T_{\mathrm{ave}}(q, n, k)\geq n - \frac{n(n - k)}{k}(H_n - H_{n-k}) +  \min_{1\leq r\leq k}&\left\{S(r)\right\}.
		\end{align*}
	\end{theorem}	
	\begin{IEEEproof}[Proof of Theorem \ref{thm:lower bound k + 1}]
		Let $G\in\bbF_q^{k\times n}$ be a rank-$k$ generator matrix, and let $r$ be the smallest positive integer such that some $r$-dimensional subspace contains $r + 1$ columns of $G$. We analyze $\sum_{i=1}^k\alpha_i^s(G)$ for $s\leq k$ as follows:
		\begin{itemize}
			\item When $s < r$: Since the subspaces generated by distinct $s$-element sets are pairwise different, each $s$-dimensional standard subspace contains columns from at most one $s$-element set. Thus, $\sum_{i=1}^k\alpha_i^s(G)\leq (s - 1)\binom{n}{s} + \binom{k}{s}$.
			\item When $r\leq s\leq k$: By Lemma \ref{lem:count}, it holds that $\sum_{i=1}^k\alpha_i^s(G)\leq s\binom{n}{s} - \binom{n-(r + 1)}{s - (r + 1)} - \sum_{t = 1}^{r - 1}\binom{r}{t - 1}\binom{k - r}{s - t}$.
			\item When $s \geq k + 1$: It holds that $\sum_{i=1}^k\alpha_i^s(G)\leq k\binom{n}{s}$.
		\end{itemize}
		
		By Lemma \ref{lem:alpha}, we have
		\begin{align*}
			T_{\mathrm{ave}}(G) = &nH_n - \sum_{s = 1}^{n - 1}\frac{\sum_{i=1}^k\alpha_i^s(G)}{k\binom{n-1}{s}}\\
			\geq &nH_n - \sum_{s = 1}^{r - 1}\frac{(s - 1)\binom{n}{s} + \binom{k}{s}}{k\binom{n - 1}{s}}- \sum_{s = k + 1}^{n - 1}\frac{k\binom{n}{s}}{k\binom{n-1}{s}} - \sum_{s = r}^{k}\frac{s\binom{n}{s} - \binom{n-(r + 1)}{s - (r + 1)} - \sum_{t = 1}^{r - 1}\binom{r}{t - 1}\binom{k - r}{s - t}}{k\binom{n-1}{s}} \\
			= &n - \frac{n(n - k)}{k}(H_n - H_{n-k}) + \sum_{s = 1}^{r - 1}\frac{\binom{n}{s} - \binom{k}{s}}{k\binom{n - 1}{s}} + \sum_{s = r}^k\frac{\binom{n - (r + 1)}{s - (r + 1)} + \sum_{t = 1}^{r - 1}\binom{r}{t - 1}\binom{k - r}{s - t}}{k\binom{n - 1}{s}}.
		\end{align*}
		
		Combining these results, we conclude that $T_{\mathrm{ave}}(q, n, k)$ is bounded below by
		\begin{align*}
			n - \frac{n(n - k)}{k}(H_n - H_{n-k}) + \min_{1\leq r\leq k}&\left\{S(r)\right\}.
		\end{align*}
	\end{IEEEproof}
	Corollary \ref{cor:n=k+1} determines $T_{\max}(q, n, k)$ when $n = k + 1$.
	\begin{corollary}\label{cor:n=k+1}
		For any $q, n, k$ with $n = k + 1$, it holds that
		$$T_{\max}(q, n = k + 1, k) = T_{\mathrm{ave}}(q, n = k + 1, k) = k.$$
	\end{corollary}	
	\begin{IEEEproof}[Proof of Corollary \ref{cor:n=k+1}]
		When $n = k + 1$, consider the generator matrix $G_r = [\be_1, \ldots, \be_k, \be_1+\ldots+\be_r]$. 
		It is clear that any $s\leq r$ columns are linearly independent. When $r\leq s\leq k$, let $U$ be a $s$-element set such that $\langle U\rangle$ is not a $s$-dimensional standard space. Then one of the following conditions must hold:
		\begin{itemize}
			\item $|U\cap\{\be_1,\ldots, \be_r, \be_1+\ldots+\be_r\}| = r + 1$. The number of such sets is $\binom{n - (r + 1)}{s - (r + 1)}$.
			\item $|U\cap\{\be_1,\ldots, \be_r, \be_1+\ldots+\be_r\}| = t$ for some $t < r$ and $\be_1+\ldots+\be_r\in U$. The number of such sets is $\binom{r}{t-1}\binom{k-r}{s-t}$.
		\end{itemize}
		In total, it can be seen that 
		\begin{align*}
			\zeta^s(G_r) &= \binom{n-(r+1)}{s-(r+1)} + \sum_{t = 1}^{r-1}\binom{r}{t-1}\binom{k-r}{s-t} =\binom{k}{s - 1} - r\binom{k - r}{s - r}.
		\end{align*}
		
		Furthermore, for each $s$-element set $U$ such that $\langle U\rangle$ is not a $s$-dimensional standard space, it contains $s-1$ standard basis vectors. This fact implies that
		\begin{align*}
			T_{\mathrm{ave}}(G) = &nH_n - \sum_{s = 1}^{k}\frac{\sum_{i=1}^k\alpha_i^s(G)}{k\binom{n-1}{s}}\\
			= &nH_n - \sum_{s = 1}^{r - 1}\frac{(s - 1)\binom{n}{s} + \binom{k}{s}}{k\binom{n - 1}{s}} + \sum_{s = r}^{k}\frac{s\left(\binom{n}{s} - \zeta^s(G_r)\right) + (s - 1)\zeta^s(G_r)}{k\binom{n - 1}{s}}\\
			= &n - \frac{n(n - k)}{k}(H_n - H_{n-k}) + S(r)\\
			= &k + 1 - \frac{k + 1}{k}(H_{k + 1} - 1) + S(r).
		\end{align*}

		When $n = k + 1$, we have
		\begin{align*}
			S(r) & = \sum_{s = 1}^{r - 1}\frac{\binom{k + 1}{s} - \binom{k}{s}}{k\binom{k}{s}} + \sum_{s = r}^k\frac{\binom{k - r}{s - (r + 1)} + \sum\limits_{t = 1}^{r - 1}\binom{r}{t - 1}\binom{k - r}{s - t}}{k\binom{k}{s}}\\
			&= \sum_{s = 1}^{r - 1}\frac{s}{k(k + 1- s)} + \sum_{s = r}^k\frac{\binom{k}{s - 1} - r\binom{k - r}{s - r}}{k\binom{k}{s}}\\
			&= \sum_{s = 1}^{r - 1}\frac{s}{k(k + 1- s)} + \sum_{s = r}^k\frac{s}{k(k + 1- s)} - \sum_{s = r}^k\frac{r\binom{k - r}{s - r}}{k\binom{k}{s}}\\
			&= \sum_{s = 1}^{k}\frac{s}{k(k + 1- s)} - \frac{r}{k}\sum_{s = r}^k\frac{\binom{k - r}{s - r}}{\binom{k}{s}}\\
			&= \frac{k+1}{k}H_k - 1 - \frac{r}{k}\sum_{s = r}^k\frac{\binom{s}{r}}{\binom{k}{r}}\\
			&= \frac{k+1}{k}H_k - 1 - \frac{r}{k}\frac{\binom{k + 1}{r + 1}}{\binom{k}{r}}\\
			&= \frac{k+1}{k}H_k - 1 - \frac{r(k + 1)}{k(r + 1)}.
		\end{align*}

		In total, we have
		\[
			T_{\mathrm{ave}}(G_r) = k + 1 - \frac{r(k + 1)}{k(r + 1)},
		\]
		which yields $T_{\mathrm{ave}}(q, n = k + 1, k) = k$.
	\end{IEEEproof}
	\begin{remark}
		When $n = k + 1$ and $k$ is sufficiently large, the lower bounds from \cite{daniella2023cover} yield
		\begin{align*}
			T_{\max}(q, n, k)&\geq \max\left\{\frac{k + 1}{2}, n - \frac{n(n - k)}{k}(H_n - H_{n - k})\right\} = k - \frac{k + 1}{k}\log k + o(\log k).
		\end{align*}
		Thus, Theorem \ref{thm:lower bound k + 1} improves the lower bounds of \cite{daniella2023cover} by $\Omega(\log k)$ for $n = k + 1$.
	\end{remark}
	Table \ref{tab:lower_bounds_comparison} compares the lower bounds on $T_{\mathrm{ave}}(q, n, k)$ between Theorem~\ref{thm:lower bound k + 1} and prior work~\cite{daniella2023cover} for $k=100$ and $n = k + c$ ($c\in [1,5]$).
	\begin{table}[h]
		\centering		\caption{Comparison of the lower bounds on $T_{\mathrm{ave}}(q, n, k)$ between Theorem~\ref{thm:lower bound k + 1} and prior work~\cite{daniella2023cover} for $k=100$ and $n = k + c$ ($c\in [1,5]$).}\label{tab:lower_bounds_comparison}
		\begin{tabular}{|c|c|c|c|c|c|}
			\hline
			$n$ & $k + 1$ & $k + 2$ & $k + 3$ & $k + 4$ & $k + 5$ \\ \hline
			Theorem \ref{thm:lower bound k + 1} & $100.00$ & $95.28$ & $93.03$ & $91.28$ & $89.79$ \\ \hline
			Lower bounds in \cite{daniella2023cover} & $96.76$  & $94.44$ & $92.55$ & $90.92$ & $89.50$ \\ \hline
		\end{tabular}
	\end{table}		
	\begin{example}
		Let $k = 3$ and $n = 4$. For any generator matrix $G\in \bbF_q^{k\times n}$, the lower bounds in \cite{daniella2023cover} imply that
		\begin{align*}
			T_{\max}(G) \geq \max\left\{\frac{k + 1}{2}, n - \frac{n(n - k)}{k}(H_n - H_{n - k})\right\} = \max\left\{2, \frac{23}{9}\right\} = \frac{23}{9}.
		\end{align*}
		
		Theorem \ref{thm:lower bound k + 1} derives the tight bound as follows:
		$$T_{\max}(G)\geq T_{\mathrm{ave}}(G)\geq \min\left\{\frac{10}{3}, \frac{28}{9}, 3\right\} = 3 > \frac{23}{9}.$$ 
	\end{example}
	Next we analyze the asymptotic behavior of the lower bound in Theorem \ref{thm:lower bound k + 1} for larger $n$. We start with the case $n = k + c$ for some constant $c$.
	\begin{corollary}\label{cor:constant_gap}
		For $n = k + c$ with constant $c$, when $k\to\infty$, it holds that
		$$T_{\mathrm{ave}}(q, n, k) \geq n - \frac{n(n - k)}{k}(H_n - H_{n-k}) + \ln 3 - \ln 2.$$
    \end{corollary}
	The following combinatorial equality is crucial in the proof of Corollary \ref{cor:constant_gap}.
	\begin{lemma}\label{lem:combinatorial_equality_2}
		For any integers $n\geq 1$ and $0\leq a\leq n$, it holds that 
		$$\sum_{s = 0}^{n - 1}\frac{\binom{n - a}{s - a}}{\binom{n - 1}{s}} = n\left(H_n - H_a\right).$$
	\end{lemma}
	\begin{IEEEproof}[Proof of Lemma \ref{lem:combinatorial_equality_2}]
	For simplicity, let $F(n, a) = \sum_{s = 0}^{n - 1}\frac{\binom{n - a}{s - a}}{\binom{n - 1}{s}}$. Note that 
	$$\frac{1}{\binom{n - 1}{s}} = \frac{s!(n - 1 - s)!}{(n - 1)!} = \frac{n}{n + 1}\left(\frac{s!(n-s)!}{n!} + \frac{(s + 1)!(n - s - 1)!}{n!}\right) = \frac{n}{n + 1}\left(\frac{1}{\binom{n}{s}} + \frac{1}{\binom{n}{s + 1}}\right),$$
	it follows that
	\begin{align*}
		F(n, a) &= \frac{n}{n + 1}\sum_{s = 0}^{n - 1}\binom{n - a}{s - a}\left(\frac{1}{\binom{n}{s}} + \frac{1}{\binom{n}{s + 1}}\right) \\
		&= \frac{n}{n + 1}\left(\sum_{s = 0}^{n - 1}\frac{\binom{n - a}{s - a}}{\binom{n}{s}} + \sum_{s = 0}^{n - 1}\frac{\binom{n - a}{s - a}}{\binom{n}{s + 1}}\right)\\
		&= \frac{n}{n + 1}\left(\sum_{s = 0}^{n}\frac{\binom{n - a}{s - a}}{\binom{n}{s}} - 1 + \sum_{s = 0}^{n}\frac{\binom{n - a}{s - a - 1}}{\binom{n}{s}}\right)\\
		&= \frac{n}{n + 1}\left(\sum_{s = 0}^{n}\frac{\binom{n - a}{s - a} + \binom{n - a}{s - a - 1}}{\binom{n}{s}} - 1\right)\\
		&= \frac{n}{n + 1}(F(n + 1, a) - 1).
	\end{align*}
	Consequently, we have $F(n + 1, a) = \frac{n + 1}{n}F(n, a) + 1$. Since $F(a, a) = 0 = a\left(H_a - H_a\right)$ and
	$$(n + 1)\left(H_{n+1} - H_a\right) = (n + 1)\left(H_n - H_a\right) + 1 = \frac{n + 1}{n}\left(n\left(H_n - H_a\right)\right) + 1,$$
	it follows that $F(n, a) = n\left(H_n - H_a\right)$ for any $n\geq a$.
	\end{IEEEproof}
	Now we are ready to prove Corollary \ref{cor:constant_gap}.
	\begin{IEEEproof}[Proof of Corollary \ref{cor:constant_gap}]
		Let $n = k + c$ for $c \geq 2$. When $r\geq \frac{2k}{3} + 1$, it holds that
		\begin{align*}
			S(r) &\geq \sum_{s=1}^{r-1}\frac{\binom{n}{s}-\binom{k}{s}}{k\binom{n-1}{s}}\\ &= \sum_{s=1}^{r-1}\frac{n}{k(n - s)} - \sum_{s=1}^{r - 1}\frac{\binom{k}{s}}{k\binom{n-1}{s}} \\
			& = \frac{n}{k}(H_{n-1} - H_{n - r}) - \sum_{s=1}^{r - 1}\frac{\binom{k}{s}}{k\binom{n-1}{s}}\\
			&\geq \frac{n}{k}(H_{n-1} - H_{n - r}) - \sum_{s=1}^{r - 1}\frac{1}{k}.
		\end{align*}
		It can be seen that $\frac{n}{k}(H_{n-1} - H_{n - r}) - \sum_{s=1}^{r - 1}\frac{1}{k}$ is increasing with $r$. Thus, when $r \geq \frac{2k}{3} + 1$, we have
		\begin{align*}
			S(r) & \geq \frac{n}{k}(H_{n-1} - H_{n - \lfloor\frac{2k}{3} + 1\rfloor}) - \frac{\lfloor\frac{2k}{3}\rfloor}{k} \gtrsim \ln 3 - \frac{2}{3}.
		\end{align*}

		When $r \leq \frac{2k}{3}$, by Lemma \ref{lem:combinatorial_equality_2}, it holds that
		\begin{align*}
			S(r) & \geq \sum_{s = r}^k\frac{\binom{n - r - 1}{s - r - 1}}{k\binom{n-1}{s}} = \frac{n}{k}(H_n - H_{r + 1}) - \sum_{s = k + 1}^{n - 1}\frac{\binom{n - r - 1}{s - r - 1}}{k\binom{n-1}{s}} \geq \frac{n}{k}(H_n - H_{r + 1}) - \sum_{s = k + 1}^{n - 1}\frac{1}{k} \gtrsim \ln \frac{3}{2}.
		\end{align*}
		As $\ln 3 - \frac{2}{3} > \ln \frac{3}{2}$, it holds that
		\begin{align*}
			T_{\mathrm{ave}}(q, n, k)\geq &n - \frac{n(n - k)}{k}(H_n - H_{n-k}) + \ln \frac{3}{2}.
		\end{align*}
	\end{IEEEproof}
	\begin{remark}
		For $n\geq k\geq 1$, it holds that
		\begin{align*}
			n - \frac{n(n - k)}{k}(H_n - H_{n-k}) &= n - \frac{n(n - k)}{k}\sum_{i = 0}^{k - 1}\frac{1}{n - i}\\
			&= n - \frac{(n - k)}{k}\sum_{i = 0}^{k - 1}\frac{n}{n - i}\\
			&= n - \frac{(n - k)}{k}\left(k + \sum_{i = 0}^{k - 1}\frac{i}{n - i}\right)\\
			&= k - \sum_{i = 0}^{k - 1}\frac{i}{k}\frac{n - k}{n - i}\\
			&\geq k - \sum_{i = 0}^{k - 1}\frac{i}{k}\\
			&= \frac{k + 1}{2}.
		\end{align*}
		Therefore, Theorem \ref{thm:lower bound k + 1} improves upon the bounds in \cite{daniella2023cover} by at least a constant additive gap.
	\end{remark}
	Finally, we show a simplified lower bound on $T_{\mathrm{ave}}(q, n, k)$ for any $n\geq k + 1$.
	\begin{corollary}\label{cor:lower bound}
		For any $q, n, k$ with $n\geq k + 1$, it holds that 
		\begin{align*}
			T_{\mathrm{ave}}(q, n, k)\geq &\,n - \frac{n(n - k)}{k}(H_n - H_{n-k}) + \min\left\{\frac{k^2-1}{3 (n-1)(n-2)},\ \frac{n-k}{k(n-1)}\right\}.
		\end{align*}
	\end{corollary}
	\begin{IEEEproof}[Proof of Corollary \ref{cor:lower bound}]
		We have the following bounds hold:
		\begin{itemize}
			\item When $r = 1$, since $S(1)=\frac{1}{k}\sum_{s=2}^{k}\frac{\binom{n-2}{s-2}}{\binom{n-1}{s}}$ and $\frac{\binom{n-2}{s-2}}{\binom{n-1}{s}}=\frac{s(s-1)}{(n-1)(n-s)}$ with $n-s\le n-2$ for $s\ge 2$, we have 
			\begin{align*}
				S(1) &=\frac{1}{k}\sum_{s=2}^k\frac{s(s-1)}{(n-1)(n-s)}\geq \frac{1}{k}\sum_{s=2}^k\frac{s(s-1)}{(n-1)(n-2)}\\
				&=\frac{1}{k}\cdot\frac{k(k+1)(k-1)}{3(n-1)(n-2)}
				=\frac{k^2-1}{3(n-1)(n-2)}.
			\end{align*}
			\item When $r \geq 2$, the first sum in $S(r)$ contains the $s=1$ term, and all remaining terms are nonnegative; it follows that
			$$S(r)\ge \frac{\binom{n}{1}-\binom{k}{1}}{k\binom{n-1}{1}}
			=\frac{n-k}{k(n-1)}.$$
		\end{itemize}
		Combining these results, we have
		\begin{align*}
			T_{\mathrm{ave}}(q, n, k)\geq &n - \frac{n(n - k)}{k}(H_n - H_{n-k}) + \min\left\{\frac{k^2-1}{3 (n-1)(n-2)},\ \frac{n-k}{k(n-1)}\right\}.
		\end{align*}
	\end{IEEEproof}
	To the best of our knowledge, the lower bound in Theorem \ref{thm:lower bound k + 1} is not tight for $n\geq k + 2$. Further investigation of this case is left for future work.

\section{Expected Number of Samples Required to Recover All Information Strands}\label{sec:expectation_number_of_all_samples}

In this section, we focus on deriving bounds on $T(q, n, k)$. When $n = k + 1$ or $q\geq n - 1$, there exists an MDS code with generator matrix $G\in \bbF_q^{k\times n}$, which implies that $T(q, n, k) = n(H_{n} - H_{n - k})$. In the following part, we focus on the case when $n \geq k + 2$. Recall that when $s\geq k$, $\zeta^s(G)$ is the number of $s$-element column sets of $G$ whose $\mathbb{F}_q$-span is not $\bbF_q^k$. We start by analyzing the behavior of $\zeta^k(G)$ for matrices $G\in\bbF_q^{k\times (k + 2)}$.

\begin{lemma}\label{lem:zeta_lower_bound_k+2}
    Given any matrix $G\in\bbF_q^{k\times (k + 2)}$ such that $q \leq k + 1$, it holds that 
    $$\zeta^k(G)\geq \frac{(k + 2)^2 - (k + 2)(q + 1)}{2(q + 1)}.$$
\end{lemma}
\begin{IEEEproof}[Proof of Lemma \ref{lem:zeta_lower_bound_k+2}]
    If the rank of $G$ is less than $k$, then $\zeta^k(G) = \binom{k + 2}{k}$, which satisfies the desired bound. Thus, we assume that $G$ has full row rank. Without loss of generality, let $G = [\be_1,\ldots,\be_k, \bu, \bv]$, where $\bu = \sum_{i = 1}^ku_i\be_i$ and $\bv = \sum_{i = 1}^kv_i\be_i$.
    Consider a sub-matrix $G'$ of size $k\times k$ of $G$. If $G'$ contains neither $\bu$ nor $\bv$, then $G'$ is always non-singular. We consider the following cases:
    \begin{itemize}
        \item $G'$ contains $\bu$ but not $\bv$. Without loss of generality, let $\be_i$ be not selected. It can be seen that $G'$ is singular if and only if $u_i = 0$.
        \item $G'$ contains $\bv$ but not $\bu$. Without loss of generality, let $\be_i$ be not selected. It can be seen that $G'$ is singular if and only if $v_i = 0$.
        \item $G'$ contains both $\bu$ and $\bv$. Without loss of generality, let $\be_i, \be_j$ be not selected. It can be seen that $G'$ is singular if and only if $u_iv_j - u_jv_i = 0$, i.e. $(u_i, v_i)$ and $(u_j, v_j)$ are linearly dependent.
    \end{itemize}
    Let $n_{(a, b)}$ be the number of $(u_i, v_i)$ contained in $\{(ta, tb): t\in \bbF_q\backslash\{0\}\}$, where 
    $$(a, b)\in S:= \{(0, 0), (0, 1), (1, 0)\}\cup\{(1, b):b\in\bbF_q\backslash\{0\} \}.$$ 
    It can be seen that
    \begin{align*}
        \zeta^k(G) = f(n_{(0, 0)}, n_{(0, 1)}, \ldots)&:= n_{(0, 0)} + n_{(0, 1)} + n_{(0, 0)} + n_{(1, 0)} + n_{(0, 0)}(k - n_{(0, 0)}) + \sum_{(a, b)\in S}\binom{n_{(a, b)}}{2}.
    \end{align*}
    Now we consider the minimum of $f$. We claim that the minimum is attained when $n_{(0, 0)} = 0$. Suppose that $n_{(0, 0)}\geq 1$, we have
    \begin{align*}
        f(n_{(0, 0)}, n_{(0, 1)}, n_{(1, 0)}, n_{(1, 1)}, \ldots) - f(n_{(0, 0)} - 1, n_{(0, 1)}, n_{(1, 0)}, n_{(1, 1)} + 1, \ldots) = 2 + k - n_{(0, 0)} - n_{(1, 1)} > 0,
    \end{align*}
    which completes the proof of the claim. Therefore, we can assume that $n_{(0, 0)} = 0$, and
    \begin{align*}
        \zeta^k(G) = f(0, n_{(0, 1)}, n_{(1, 0)}, n_{(1, 1)}, \ldots):= &n_{(0, 1)} + n_{(1, 0)} + \sum_{(a, b)\in S}\binom{n_{(a, b)}}{2}\\
        =&\binom{n_{(0, 1)} + 1}{2} + \binom{n_{(1, 0)} + 1}{2} + \sum_{(a, b)\in S\backslash\{(0, 0), (0, 1), (1, 0)\}}\binom{n_{(a, b)}}{2}.
    \end{align*}
    As $\binom{x}{2}$ is convex, by discrete Jensen's inequality, we have
    \begin{align*}
        \zeta^k(G) &\geq (q + 1 - R)\binom{Q}{2} + R\binom{Q + 1}{2} \geq (q + 1)\binom{\frac{k + 2}{q + 1}}{2} = \frac{(k + 2)^2 - (k + 2)(q + 1)}{2(q + 1)},
    \end{align*}
    where $k + 2 = Q(q + 1) + R$ and $0\leq R< q$, i.e. $Q = \lfloor\frac{k + 2}{q + 1}\rfloor$ and $R = (k + 2) - (q + 1)\lfloor\frac{k + 2}{q + 1}\rfloor$.
\end{IEEEproof}

With the lower bound on $\zeta^k(G)$, we can derive the following lower bound on $T(q, n, k)$ for $n\geq k + 2$.
\begin{theorem}\label{thm:whole_lower_bound}
	For any $q, n, k$ with $n\geq k + 2$ and $q \leq k + 1$, it holds that
	\begin{align*}
		T(q, n, k) &\geq n(H_n - H_{n-k}) + \frac{(k - q + 1)n}{(k + 1)(q + 1)(n-k)}.
	\end{align*}
\end{theorem}
\begin{IEEEproof}[Proof of Theorem \ref{thm:whole_lower_bound}]
	Let $G\in\bbF_q^{k\times n}$ be a generator matrix and $G'$ be a $k\times (k + 2)$ sub-matrix of $G$. By Lemma \ref{lem:zeta_lower_bound_k+2}, we have
	\begin{align*}
		\zeta^k(G')&\geq \frac{(k + 2)^2 - (k + 2)(q + 1)}{2(q + 1)} = \frac{k - q + 1}{(k + 1)(q + 1)}\binom{k + 2}{2}.
	\end{align*}
	As there are $\binom{n}{k + 2}$ such sub-matrices, and each $k\times k$ sub-matrix is contained in $\binom{n - k}{2}$ such $k\times (k + 2)$ sub-matrices, it follows that
	\begin{align*}
		\zeta^k(G) &\geq \frac{\binom{n}{k + 2}}{\binom{n - k}{2}}\frac{k - q + 1}{(k + 1)(q + 1)}\binom{k + 2}{2} = \frac{(k - q + 1)}{(k + 1)(q + 1)}\binom{n}{k}.
	\end{align*}
	By Lemma \ref{lem:whole_alpha}, it holds that
	\begin{align*}
		\bbE[\tau(G)] &= nH_n - \sum_{s = k}^{n - 1}\frac{\alpha^s(G)}{\binom{n - 1}{s}}\\
		&= nH_n - \sum_{s = k}^{n - 1}\frac{\binom{n}{s} - \zeta^s(G)}{\binom{n - 1}{s}} \\
		&\geq nH_n - \sum_{s = k}^{n -1}\frac{\binom{n}{s}}{\binom{n - 1}{s}} + \frac{(k - q + 1)\binom{n}{k}}{(k + 1)(q + 1)\binom{n - 1}{k}}\\ 
		&= n(H_n - H_{n - k}) + \frac{(k - q + 1)n}{(k + 1)(q + 1)(n-k)}.
	\end{align*}
\end{IEEEproof}
Next, we derive an upper bound on $T(q, n, k)$ by the random code construction. The key step is to analyze the probability that a $k\times s$ sub-matrix of a random matrix $G\in\bbF_q^{k\times n}$ is singular.
\begin{theorem}\label{thm:whole_upper_bound}
	For any $q, n, k$ with $n\geq k + 2$, there exists a generator matrix $G\in\bbF_q^{k\times n}$ such that
	\begin{align*}
		T(q, n, k) &\leq n(H_n - H_{n-k}) + n\sum_{t=0}^{n-k-1}\left(\frac{q^{-t}}{q-1}\right)\frac{1}{n-k-t},
	\end{align*}
	where $t = s - k$.
\end{theorem}
\begin{IEEEproof}[Proof of Theorem \ref{thm:whole_upper_bound}]
	Consider a random matrix $G\in\bbF_q^{k\times n}$, where each entry is independently and uniformly drawn from $\bbF_q$. It is clear that $\bbE_{G}[\bbE[\tau(G)]]$ is an upper bound on $T(q, n, k)$. For a $k\times s$ sub-matrix $G'$ of $G$, $G'$ is singular if and only if the rows of $G'$ are linearly dependent. It is known that the probability that $s$ randomly chosen vectors in $\bbF_q^k$ are linearly independent is given by $\prod_{i = 0}^{k - 1}(1 - q^{i-s})$. Thus, we have
	\begin{align*}
		\sum_{s=k}^{n-1}\frac{\bbE_{G}[\zeta^s(G)]}{\binom{n - 1}{s}} &= \sum_{s=k}^{n-1}\frac{\bbE_{G}[\zeta^s(G)]}{\binom{n}{s}}\frac{\binom{n}{s}}{\binom{n - 1}{s}}\\
		&= \sum_{s=k}^{n-1}\left(1 - \prod_{i = 0}^{k - 1}(1 - q^{i-s})\right)\frac{n}{n-s}\\
		&\leq \sum_{s=k}^{n-1}\left(1 - \left(1 - q^{-s}\sum_{i = 0}^{k - 1}q^{i}\right)\right)\frac{n}{n-s}\\ 
		&= \sum_{s=k}^{n-1}\left(q^{-s}\frac{q^{k}-1}{q-1}\right)\frac{n}{n-s}\\
        &= \sum_{s=k}^{n-1}\left(\frac{q^{k-s}}{q-1} - \frac{q^{-s}}{q-1}\right)\frac{n}{n-s},
	\end{align*}
	which implies that there exists a generator matrix $G\in\bbF_q^{k\times n}$ such that
	\begin{align*}
		\sum_{s=k}^{n-1}\frac{\zeta^s(G)}{\binom{n - 1}{s}} &\leq \sum_{s=k}^{n-1}\left(\frac{q^{k-s}}{q-1} - \frac{q^{-s}}{q-1}\right)\frac{n}{n-s}.
	\end{align*}
	By Lemma \ref{lem:whole_alpha}, we have
    \begin{align*}
        \bbE[\tau(G)] - n(H_n-H_{n-k})& = \sum_{s=k}^{n-1} \frac{\binom{n}{s} - \alpha^s(G)}{\binom{n-1}{s}} = \sum_{s=k}^{n-1} \frac{\zeta^s(G)}{\binom{n-1}{s}},
    \end{align*}
    which implies that
    \begin{align*}
        \bbE[\tau(G)] - n(H_n-H_{n-k})&\leq \sum_{s=k}^{n-1}\left(\frac{q^{k-s}}{q-1} - \frac{q^{-s}}{q-1}\right)\frac{n}{n-s}\\
        &\leq \sum_{s=k}^{n-1}\left(\frac{q^{k-s}}{q-1}\right)\frac{n}{n-s}\\
        & = n\sum_{t=0}^{n-k-1}\left(\frac{q^{-t}}{q-1}\right)\frac{1}{n-k-t},
    \end{align*}
    where $t = s - k$.
\end{IEEEproof}

With Theorem \ref{thm:whole_lower_bound} and Theorem \ref{thm:whole_upper_bound}, we can derive the following asymptotic result on $T(q, n, k)$ when $q$ is fixed and $k = Rn$ for some constant $R < 1$.

\begin{corollary}\label{cor:whole_asymptotic_Rn}
    For any fixed $q$ and $k = Rn$ for some constant $R < 1$, it holds that
    \begin{align*}
        T(q, n, k = Rn) = n(H_n - H_{n-k}) + \Theta(1)~\text{and}~\frac{T(q, n, k = Rn)}{k} \to \frac{1}{R}\ln \frac{1}{1 - R} ~~(k\to\infty).
    \end{align*}
\end{corollary}
\begin{IEEEproof}[Proof of Corollary \ref{cor:whole_asymptotic_Rn}]
We first derive the lower bound and upper bound on $T(q, n, k) - n(H_n - H_{n-k})$ when $q$ is fixed and $k = Rn$ for some constant $R < 1$. By Theorem \ref{thm:whole_lower_bound}, it holds that
	\begin{align*}
		T(q, n, k = Rn) - n(H_n - H_{n-k}) &\geq  \frac{(k - q + 1)n}{(k + 1)(q + 1)(n-k)}\\
		&= \frac{1}{(q + 1)(1 - R)} + o(1)~~(k\to\infty).
	\end{align*}

    By Theorem \ref{thm:whole_upper_bound}, we have
	\begin{align*}
		T(q, n, k = Rn) - n(H_n - H_{n-k}) &\leq n\sum_{t=0}^{n-k-1}\left(\frac{q^{-t}}{q-1}\right)\frac{1}{n-k-t} = n\sum_{t=0}^{(1-R)n-1}\left(\frac{q^{-t}}{q-1}\right)\frac{1}{(1 - R)n-t}.
	\end{align*}

	When $t\leq \left\lfloor\frac{(1-R)n}{2}\right\rfloor$, it holds that $(1-R)n-t\geq \frac{(1-R)n}{2}$, which implies that
    \begin{align*}
        \sum_{t=0}^{\left\lfloor\frac{(1-R)n}{2}\right\rfloor}\left(\frac{q^{-t}}{q-1}\right)\frac{n}{(1-R)n-t}\leq \sum_{t=0}^{\left\lfloor\frac{(1-R)n}{2}\right\rfloor}\left(\frac{q^{-t}}{q-1}\right)\frac{2n}{(1-R)n}\leq \frac{2q}{(1-R)(q - 1)^2}.
    \end{align*}
    When $t> \left\lfloor\frac{(1-R)n}{2}\right\rfloor$, we have 
    \begin{align*}
        \sum_{t=\left\lfloor\frac{(1-R)n}{2}\right\rfloor + 1}^{(1-R)n-1}\left(\frac{q^{-t}}{q-1}\right)\frac{n}{(1-R)n-t}\leq \sum_{t=\left\lfloor\frac{(1-R)n}{2}\right\rfloor + 1}^{(1-R)n-1}\left(\frac{q^{-t}}{q-1}\right)n \leq \frac{q^{-\left\lfloor\frac{(1-R)n}{2}\right\rfloor}}{(q - 1)^2}n.
    \end{align*}
    In total, we have 
    $$T(q, n, k = Rn) - n(H_n-H_{n-k})\leq \frac{2q}{(1-R)(q - 1)^2} + \frac{q^{-\left\lfloor\frac{(1-R)n}{2}\right\rfloor}}{(q - 1)^2}n\to \frac{2q}{(1-R)(q - 1)^2}~~(k\to\infty).$$

	In total, when $q$ is fixed and $k = Rn$ for some constant $R < 1$, it holds that
	\begin{align*}
		\frac{1}{(q + 1)(1 - R)}\leq T(q, n, k = Rn) - n(H_n - H_{n-k}) \leq \frac{2q}{(1-R)(q - 1)^2} ~~(k\to\infty).
	\end{align*}
	As $H_n = \ln n + \gamma + o(1)$ as $n \to \infty$, the term $\frac{T(q, n, k = Rn)}{k}$ converges to $\frac{1}{R}\ln \frac{1}{1 - R}$ as $k \to \infty$.
\end{IEEEproof}

An interesting open problem is to determine the exact value of $\lim_{k\to\infty}T(q, n, k) - n(H_n - H_{n-k})$ when $q$ is fixed and $k = Rn$ for some constant $R < 1$, which is left for future work. We close this section by extending the asymptotic result in Corollary \ref{cor:whole_asymptotic_Rn} to the case where $n - k$ is either a constant or $n - k\to\infty$.

\begin{theorem}\label{thm:whole_asymptotic}
    For any fixed $q$, assume that $n - k = f(n)$ with $f(n)$ is either a constant or $f(n) \to \infty$, it holds that
    \begin{align*}
        T(q, n, k) = n(H_n - H_{n-k}) + \Theta\left(\frac{n}{f(n)}\right) ~~(k\to\infty).
    \end{align*}
\end{theorem}
\begin{IEEEproof}[Proof of Theorem \ref{thm:whole_asymptotic}]
    By Theorem \ref{thm:whole_lower_bound}, we have
    \begin{align*}        
        T(q, n, k) &\geq n(H_n - H_{n-k}) + \frac{(k - q + 1)n}{(k + 1)(q + 1)(n-k)}\\
        & = n(H_n - H_{n-k}) + \Theta\left(\frac{n}{f(n)}\right) ~~(k\to\infty).
    \end{align*}

    By Theorem \ref{thm:whole_upper_bound}, we have
    \begin{align*}
        T(q, n, k) \leq n(H_n - H_{n-k}) + n\sum_{t=0}^{n-k-1}\left(\frac{q^{-t}}{q-1}\right)\frac{1}{n-k-t}.
    \end{align*}
    
    If $f(n)$ is a constant, we have 
    \begin{align*}
        n\sum_{t=0}^{f(n)-1}\left(\frac{q^{-t}}{q-1}\right)\frac{1}{f(n)-t} = \Theta(n)~~(k\to\infty),
    \end{align*}
    which implies that $T(q, n, k) - n(H_n - H_{n-k}) = \Theta(n)$ when $f(n)$ is a constant. Furthermore, if $f(n)\to\infty$, it holds that
    \begin{align*}
        \sum_{t=0}^{f(n)-1}\left(\frac{q^{-t}}{q-1}\right)\frac{1}{f(n)-t} = &\frac{q^{-f(n)}}{q-1}\sum_{t = 1}^{f(n)}\frac{q^t}{t} \\
        = &\frac{q^{-f(n)}}{q-1}\left(\sum_{t = 1}^{\lfloor f(n) / 2\rfloor}\frac{q^t}{t} + \sum_{t = \lfloor f(n) / 2\rfloor + 1}^{f(n)}\frac{q^t}{t}\right)\\
        \leq &\frac{q^{-f(n)}}{q-1}\left(q^{\frac{f(n)}{2}}(\ln f(n) + 1) + 2\frac{q^{f(n)}}{f(n)}\sum_{t = \lfloor f(n) / 2\rfloor + 1}^{f(n)}q^{t - f(n)}\right)\\
        \leq &\frac{q^{-f(n)}}{q-1}\left(q^{\frac{f(n)}{2}}(\ln f(n) + 1) + 2\frac{q^{f(n)}}{f(n)}\frac{q}{q - 1}\right)\\
        = &\Theta\left(\frac{1}{f(n)}\right)~~(k\to\infty).
    \end{align*}
    Thus, we have $T(q, n, k) - n(H_n - H_{n-k}) = \Theta(\frac{n}{f(n)})$ when $f(n)\to\infty$.
\end{IEEEproof}

\section{Conclusion}\label{sec:conclusion}
In this work, we addressed the random access problem in DNA storage. For the random access setting, we derived efficient algorithms for determining the expected number of samples required for recovery. Furthermore, we provided new upper bounds on $T_{\mathrm{ave}}(q, n, k)$ and $T_{\mathrm{ave}}(q, k)$ for $k = 3$ and $k = 4$. On the lower bound side, we characterized $T_{\mathrm{ave}}(q, n = k + 1, k)$ for arbitrary $q$ and $k$, and improved the lower bound on $T_{\mathrm{ave}}(q, n, k)$ for $n > k + 1$. For the non-random access setting, we derived new lower and upper bounds on $T(q, n, k)$, thereby determining its asymptotic behavior. Despite these advances, several intriguing questions remain open and warrant further investigation. We highlight a few promising directions below.
\begin{itemize}
	\item Our algorithm for computing $\bbE[\tau_i(G)]$ is efficient when $q$ and $k$ are small, but it becomes computationally intensive as $q$ and $k$ grow. It would be interesting to develop more efficient algorithms for computing $\bbE[\tau_i(G)]$ when $q$, $k$ and $n$ are large.
	\item For $k = 3$ and $n \to \infty$, we constructed a generator matrix $G$ in which all vectors of the same weight appear with equal probability. Through computer search, we observed that $T_{\mathrm{ave}}(G)$ coincides with the lower bound for small $q$. It would be worthwhile to determine whether this holds for $k = 3$ and, more generally, for $k \geq 4$.
	\item Based on the computed values of $T_{\mathrm{ave}}(q, n, k)$ and $T_{\max}(q, n, k)$ for $q = 2$, $k = 3$, and $n \in [4, 8]$, we observe that $T_{\mathrm{ave}}(q, n, k)$ does not always equal $T_{\max}(q, n, k)$. It remains an open question whether $$\liminf_{n\to\infty} T_{\mathrm{ave}}(q, n, k) = \liminf_{n\to\infty} T_{\max}(q, n, k).$$
	\item For $q = 2$, $k = 3$, and $n \in [4, 8]$, we determined the exact values of $T_{\mathrm{ave}}(q, n, k)$ and $T_{\max}(q, n, k)$ via exhaustive search. However, it is still unknown whether there exists an integer $N$ such that $T_{\mathrm{ave}}(q, n, k)$ and $T_{\max}(q, n, k)$ become monotonic in $n$ for all $n > N$.
	\item For $n > k + 1$, the gap between our lower and upper bounds on $T_{\mathrm{ave}}(q, n, k)$ remains substantial. Narrowing this gap is an important direction for future work.
	\item For $n > k + 1$ and $q \leq k$, we characterized the asymptotic behavior of $T(q, n, k)$. A natural next step is to determine the constant in $\lim_{k\to\infty}T(q, n, k) - n(H_n - H_{n-k})$ when $q$ is fixed and $k = Rn$ for some constant $R < 1$.
	\item For $n > k + 1$ and $q \leq k$, we derived an upper bound on $T(q, n, k)$ by the random code construction. It would be interesting to explore other explicit constructions to further improve the upper bound on $T(q, n, k)$ when $k = Rn$ for some constant $R < 1$.
\end{itemize}

	\section{Acknowledgments}
	The authors would like to thank Matteo Bertuzzo, Avital Boruchovsky, Anina Gruica, and Itzhak Tamo, for their helpful discussions and comments. The result of Theorem \ref{thm:whole_upper_bound} was also independently derived by Matteo Bertuzzo.

	\appendices
	\section{Proof of Theorem \ref{thm:k=4}}\label{app:k=4}
	In this part, we provide the proof of Theorem \ref{thm:k=4}. Lemma \ref{lem:weight distribution 4} is crucial for the proof of Theorem \ref{thm:k=4}.
\begin{lemma}\label{lem:weight distribution 4}
    For any rational numbers $0 \leq w_1, w_2, w_3, w_4 \leq 1$ satisfying $$4(q - 1)w_1 + 6(q - 1)^2w_2+4(q - 1)^3w_3 + (q-1)^4w_4 = 1,$$ define $x = (q - 1)w_1, y = (q - 1)w_2, z = (q - 1)w_3$, and $w = (q - 1)w_4$. Then it holds that 
    \begin{align*}
        T_{\max}(q, n, 4) \leq 1 + \frac{1}{4}&\left[(q - 1)^2(q + 1)\left(12\left(\frac{1}{1 - x} - 1\right) + 24(q-1)\left(\frac{1}{1 - y} - 1\right)\right.\right.\\ 
        &+ \left.16(q - 1)^2\left(\frac{1}{1 - z} - 1\right) + 4(q - 1)^3\left(\frac{1}{1 - w} - 1\right)\right)\\
        & + (1 - q)\left(12\left(\frac{1}{1 - 2x - (q - 1)y} - 1\right) + 36(q - 1)\left(\frac{1}{1 - x - y - (q - 1)z} - 1\right)\right.\\
        &+ 16(q - 1)^2\left(\frac{1}{1 - 3y - (q - 2)z} - 1\right) + 12(q - 1)^2\left(\frac{1}{1 - x - z - (q - 1)w} - 1\right)\\
        &+ 12(q - 1)^2\left(\frac{1}{1 - 2y - (q - 1)w} - 1\right) + 24(q - 1)^3\left(\frac{1}{1 - y - 2z - (q - 2)w} - 1\right)\\
        &\left.+ 4(q - 1)^3(q - 2)\left(\frac{1}{1 - 4z - (q - 3)w} - 1\right)\right) \\
        &+\left(4\left(\frac{1}{1 - 3x - 3(q - 1)y - (q - 1)^2z} - 1\right)\right.\\
        &+ 12(q - 1)\left(\frac{1}{1 - 2x - qy - 2(q - 1)z - (q - 1)^2w} - 1\right)\\
        &+ 12(q - 1)^2\left(\frac{1}{1 - x - 3y - (4q - 5)z - (q - 1)(q - 2)w} - 1\right)\\
        &+\left.\left.4(q - 1)^3\left(\frac{1}{1 - 6y - 4(q - 2)z - (q^2 - 3q + 3)w} - 1\right)\right)\right].
    \end{align*}
\end{lemma}
\begin{IEEEproof}[Proof of Lemma \ref{lem:weight distribution 4}]
    To prove this conclusion, we first enumerate the weight distributions and count the corresponding subspaces of $\mathbb{F}_q^4$. Table~\ref{tab:weight distribution 4} provides the number of subspaces of $\mathbb{F}_q^4$ for each weight distribution and dimension.
    \begin{table}[H]
        \centering
        \caption{Number of subspaces of $\bbF_q^4$ for each weight distribution and dimension}\label{tab:weight distribution 4}
        \begin{tabular}{|c|c|c|}
        \hline
            Dimension & Weight Distribution & Number of Subspaces \\\hline
            1 & $(q - 1, 0, 0, 0)$ & $4$ \\\hline
            1 & $(0, q - 1, 0, 0)$ & $6(q - 1)$ \\\hline
            1 & $(0, 0, q - 1, 0)$ & $4(q - 1)^{2}$ \\\hline
            1 & $(0, 0, 0, q - 1)$ & $(q - 1)^{3}$ \\\hline
            2 & $(2(q - 1), (q - 1)^2, 0, 0)$ & $6$ \\\hline
            2 & $(q - 1, q - 1, (q - 1)^2, 0)$ & $12(q - 1)$ \\\hline
            2 & $(0, 3(q - 1), (q - 1)(q - 2), 0)$ & $4(q - 1)^2$ \\\hline
            2 & $(q - 1, 0, q - 1, (q - 1)^2)$ & $4(q - 1)^2$ \\\hline
            2 & $(0, 2(q - 1), 0, (q - 1)^2)$ & $3(q - 1)^2$ \\\hline
            2 & $(0, (q - 1), 2(q - 1), (q - 1)(q - 2))$ & $6(q - 1)^3$ \\\hline
            2 & $(0, 0, 4(q - 1), (q - 1)(q - 3))$ & $(q - 1)^3(q - 2)$ \\\hline
            3 & $(3(q - 1), 3(q - 1)^2, (q - 1)^3, 0)$ & $4$ \\\hline
            3 & $(2(q - 1), q(q - 1), 2(q - 1)^2, (q - 1)^3)$ & $6(q - 1)$ \\\hline
            3 & $(q - 1, 3(q - 1), (q - 1)(4q - 5), (q - 1)^2(q - 2))$ & $4(q - 1)^2$ \\\hline
            3 & $(0, 6(q - 1), 4(q - 1)(q - 2), (q - 1)(q^2 - 3q + 3))$ & $(q - 1)^3$ \\\hline
        \end{tabular}
    \end{table}
    We first consider a $1$-dimensional subspace $\cU \leq \bbF_q^4$. Such a subspace can be written as $\cU = \langle \bu \rangle$ for some nonzero vector $\bu = h_1\be_1 + h_2\be_2 + h_3\be_3 + h_4\be_4$ with $h_i \in \bbF_q$. The number of nonzero vectors $\bu$ of Hamming weight $1$, $2$, $3$, and $4$ is $4(q-1)$, $6(q-1)^2$, $4(q-1)^3$, and $(q-1)^4$, respectively.

    We next consider a $2$-dimensional subspace $\cU \leq \bbF_q^4$. Let $\cU = \langle \bu_1, \bu_2 \rangle$, where the basis is chosen such that $\mathrm{wt}(\bu_1) \leq \mathrm{wt}(\bu_2)$ and the weight of $\bu_1, \bu_2$ is minimal among all possible bases of $\cU$. We begin with the case $\mathrm{wt}(\bu_1) = 1$.

    Recall that $\mathrm{supp}(\cU) = \bigcup_{\bu \in \cU} \mathrm{supp}(\bu)$ is the union of supports of all vectors in $\cU$. If $|\mathrm{supp}(\cU)| = 2$, then $\cU = \langle \be_i, \be_j \rangle$ for some distinct $i, j \in [4]$. There are $6$ such subspaces. Their weight distribution consists of $2(q-1)$ vectors of weight $1$, $(q-1)^2$ vectors of weight $2$, and no vectors of weight $3$ or $4$.

    If $|\mathrm{supp}(\cU)| = 3$, there are $\binom{4}{3} = 4$ possible choices for the support of $\cU$. For each such support, $\cU$ can be regarded as a $2$-dimensional subspace of $\bbF_q^3$. Applying Lemma~\ref{lem:weight distribution 3}, we obtain two types of subspaces with the following weight distributions and multiplicities:
    \begin{itemize}
    \item $(q-1, q-1, (q-1)^2, 0)$ with multiplicity $12(q-1)$;
    \item $(0, 3(q-1), (q-1)(q-2), 0)$ with multiplicity $4(q-1)^2$.
    \end{itemize}

    We now consider the case where $|\mathrm{supp}(\cU)| = 4$. Since $\mathrm{wt}(\bu_1) = 1$, we have $\bu_1 = \be_i$ for some $i \in [4]$, giving four choices for $\bu_1$. Without loss of generality, assume $\bu_1 = \be_1$ and write $\bu_2 = g_2 \be_2 + g_3 \be_3 + \be_4$ with $g_2, g_3 \in \mathbb{F}_q$. The condition $|\mathrm{supp}(\cU)| = 4$ forces $g_2$ and $g_3$ to be nonzero; otherwise, the support would be smaller. For a fixed $i$, the number of such subspaces is $(q-1)^2$, corresponding to the independent choices of $g_2, g_3 \in \mathbb{F}_q^*$. Accounting for the four choices of $i$, the total number of subspaces of this type is $4(q-1)^2$. In this subspace, the nonzero vectors are of the form $a\be_1 + b\bu_2$ with $(a,b) \neq (0,0)$. When $b = 0$ and $a \neq 0$, we obtain $(q-1)$ vectors of weight $1$. When $a = 0$ and $b \neq 0$, we obtain $(q-1)$ vectors of weight $3$. When both $a$ and $b$ are nonzero, we obtain $(q-1)^2$ vectors of weight $4$. No vectors of weight $2$ appear. Thus the weight distribution is $(q-1, 0, q-1, (q-1)^2)$.

    We now consider the case where $\mathrm{wt}(\bu_1) = 2$. Then $\mathrm{supp}(\bu_1) = \{ \be_i, \be_j \}$ for distinct $i, j \in [4]$, giving six possible choices for $\bu_1$. Without loss of generality, assume $\bu_1 = \be_1 + h_2 \be_2$ with $h_2 \neq 0$. There are two subcases.
    \begin{itemize}
        \item Subcase $\mathrm{wt}(\bu_2) = 2$: We may choose $\bu_2 = g_3 \be_3 + \be_4$ with $g_3 \neq 0$ to ensure $|\mathrm{supp}(\cU)| = 4$. Since $\langle \bu_1, \bu_2 \rangle = \langle \bu_2, \bu_1 \rangle$, the total number of such subspaces, accounting for the six choices of $\bu_1$, is $3(q-1)^2$. Their weight distribution is $(0, 2(q-1), 0, (q-1)^2)$.
        \item Subcase $\mathrm{wt}(\bu_2) = 3$: Here $\bu_2$ must be of the form $\bu_2 = g_2 \be_2 + g_3 \be_3 + \be_4$ with $g_2, g_3 \neq 0$. For each of the six choices of $\bu_1$, the number of such $\bu_2$ is $(q-1)^2$, yielding a total of $6(q-1)^3$ subspaces. Their weight distribution is $(0, q-1, 2(q-1), (q-1)(q-2))$.
    \end{itemize}
    
    We now consider the case where $\mathrm{wt}(\bu_1) = 3$. Then $\mathrm{supp}(\bu_1)$ consists of three distinct basis vectors; there are $\binom{4}{3} = 4$ possible supports. Without loss of generality, take $\mathrm{supp}(\bu_1) = \{\be_1,\be_2,\be_3\}$ and write $\bu_1 = \be_1 + h_2\be_2 + h_3\be_3$ with $h_2, h_3 \neq 0$. To ensure that the subspace has full support ${1,2,3,4}$, we choose $\bu_2 = g_2\be_2 + g_3\be_3 + \be_4$ with $g_2, g_3\neq 0$. The condition that $(\bu_1,\bu_2)$ minimizes the weight among all bases of $\cU$ forces $h_2 g_3 \neq h_3 g_2$; otherwise, $\bu_1$ and $\bu_2$ would be linearly dependent over a two-dimensional subspace with smaller support. For a fixed support of $\bu_1$, the number of admissible pairs $(h_2,h_3,g_2,g_3)$ is $(q-1)^4 - (q-1)^3 = (q-1)^3(q-2)$. Multiplying by the four possible supports yields a total of $4(q-1)^3(q-2)$ subspaces of this type. In such a subspace, the nonzero vectors consist of:
    \begin{itemize}
    \item $4(q-1)$ vectors of weight $3$ (obtained by scaling $\bu_1$ and $\bu_2$ and their linear combinations that happen to have weight $3$),
    \item $(q-1)(q-3)$ vectors of weight $4$ (the remaining nonzero combinations),
    \item no vectors of weight $1$ or $2$.
    \end{itemize}
    Hence the weight distribution is $(0,0,4(q-1), (q-1)(q-3))$.

    Finally, we consider a $3$-dimensional subspace $\cU \leq \mathbb{F}_q^4$. Let $\cH = \cU^{\bot}$ denote the orthogonal complement of $\cU$. If $\mathrm{wt}(\cH) = 1$, then $\cH = \langle \be_i \rangle$ for some $i \in [4]$, yielding four possibilities. Without loss of generality, assume $\cH = \langle \be_1 \rangle$; then $\cU$ consists of all vectors with zero first coordinate, i.e., $\cU = \langle \be_2, \be_3, \be_4 \rangle$. Consequently, the weight distribution of $\cU$ (i.e., the numbers of nonzero vectors of weight $1$, $2$, $3$, and $4$) is $(3(q-1), 3(q-1)^2, (q-1)^3, 0)$.

    If $\mathrm{wt}(\cH) = 2$, then $\cH$ is spanned by a vector of weight $2$, say $\be_i + h \be_j$ with distinct $i, j \in [4]$ and $h\neq 0$. There are $6(q-1)$ such subspaces, corresponding to the choice of the two coordinates and the nonzero scalar. Without loss of generality, assume $\cH = \langle \be_1 + h \be_2 \rangle$. Then $\cU$ consists of all vectors orthogonal to $\be_1 + h \be_2$, i.e., vectors of the form $-hg_2\be_1 + g_2\be_2+g_3\be_3+g_4\be_4$ with $g_2, g_3, g_4 \in \mathbb{F}_q$. We count the nonzero vectors in $\cU$ according to their Hamming weight.
    \begin{itemize}
        \item When $g_2 = 0$, the vector reduces to $g_3\be_3+g_4\be_4$ with $(g_3,g_4) \neq (0,0)$. This yields $2(q-1)$ vectors of weight $1$ (exactly one of $g_3,g_4$ nonzero) and $(q-1)^2$ vectors of weight $2$ (both $g_3$ and $g_4$ nonzero).
        \item When $g_2 \neq 0$, the first two coordinates are nonzero. Then: \begin{itemize}
            \item If $g_3 = g_4 = 0$, we obtain $q-1$ vectors of weight $2$ (only the first two coordinates nonzero).
            \item If exactly one of $g_3, g_4$ is nonzero, we obtain $2(q-1)^2$ vectors of weight $3$.
            \item If both $g_3$ and $g_4$ are nonzero, we obtain $(q-1)^3$ vectors of weight $4$.
        \end{itemize}
    \end{itemize}
    Summing over all possibilities, the total number of nonzero vectors in $\cU$ of each weight is $(2(q-1), q(q-1), 2(q-1)^2, (q-1)^3)$.

    If $\mathrm{wt}(\cH) = 3$, then $\cH$ is spanned by a vector of weight $3$, i.e., $\cH = \langle \be_i + h_j \be_j + h_\ell \be_\ell \rangle$ for distinct indices $i,j,\ell \in [4]$ and nonzero scalars $h_j, h_\ell \in \mathbb{F}_q^*$. The number of such subspaces is $4(q-1)^2$, corresponding to the choice of the three coordinates and the two nonzero coefficients. Without loss of generality, set $\cH = \langle \be_1 + h_2 \be_2 + h_3 \be_3 \rangle$ with $h_2, h_3 \neq 0$. Then $\cU = \cH^\perp$ consists of all vectors orthogonal to $\be_1 + h_2 \be_2 + h_3 \be_3$, i.e., vectors of the form $-(h_2g_2 + h_3g_3)\be_1 + g_2\be_2+g_3\be_3+g_4\be_4\in \mathbb{F}_q^4$ with $g_2, g_3, g_4 \in \mathbb{F}_q$. We count the nonzero vectors in $\cU$ according to their Hamming weight by analyzing the parameters.
    \begin{itemize}
        \item If $g_2 = g_3 = 0$, then $\bu = g_4\be_4$ with $g_4 \neq 0$. This case gives $(q-1)$ vectors of weight $1$.
        \item When $g_2 = 0$, $g_3\neq 0$ or $g_2 \neq 0$, $g_3 = 0$ or $g_2 \neq 0$, $g_3=-h_2g_2h_3^{-1}$, there are $q - 1$ vectors of weight $2$ and $(q - 1)^{2}$ vectors of weight $3$ for each case.
        \item When $g_2 = 0$, $g_3\neq 0, -h_2g_2h_3^{-1}$, there are $(q - 1)(q - 2)$ vectors of weight $3$ and $(q-1)^2(q - 2)$ vectors of weight $4$.
    \end{itemize}
    In total, we have $\mathrm{wd}(\cU) = ((q - 1), 3(q - 1), (q - 1)(4q - 5), (q - 1)^2(q - 2))$.

    If $\mathrm{wt}(\cH) = 4$, then $\cH$ is spanned by a vector of weight $4$, i.e., $\cH = \langle \be_1 + h_2\be_2 + h_3\be_3 + h_4\be_4 \rangle$ with $h_2, h_3, h_4 \in \mathbb{F}_q^*$. The number of such subspaces is $(q-1)^3$, corresponding to the independent choices of the three nonzero coefficients. Without loss of generality, we take this representative. The orthogonal complement $\cU = \cH^\perp$ consists of all vectors orthogonal to $\be_1 + h_2\be_2 + h_3\be_3 + h_4\be_4$, i.e., vectors of the form $-(h_2g_2 + h_3g_3 + h_4g_4)\be_1 + g_2\be_2+g_3\be_3+g_4\be_4$ with $g_2, g_3, g_4 \in \mathbb{F}_q$. We count the nonzero vectors in $\cU$ according to their Hamming weight by examining the parameters $(g_2, g_3, g_4) \neq (0,0,0)$.
    \begin{itemize}
        \item When exactly two of $g_2, g_3, g_4$ are zero. There are three possibilities: $(g_2, g_3) = (0,0)$ with $g_4 \neq 0$, $(g_2, g_4) = (0,0)$ with $g_3 \neq 0$, and $(g_3, g_4) = (0,0)$ with $g_2 \neq 0$. In each case, the first coordinate becomes a nonzero multiple of the nonzero parameter, yielding a vector with exactly two nonzero entries. Each such subcase contributes $(q-1)$ vectors, giving a total of $3(q-1)$.
        \item When exactly one of $g_2, g_3, g_4$ is zero and the first coordinate $a = -(h_2 g_2 + h_3 g_3 + h_4 g_4)$ vanishes. Consider the three subcases: $g_2 = 0$ with $g_3, g_4 \neq 0$; $g_3 = 0$ with $g_2, g_4 \neq 0$; $g_4 = 0$ with $g_2, g_3 \neq 0$. In each subcase, the condition $a = 0$ imposes a linear relation among the two nonzero parameters (e.g., $h_3 g_3 + h_4 g_4 = 0$ when $g_2 = 0$), which has exactly $q-1$ solutions. These produce vectors with the two nonzero parameters as the only nonzero coordinates, hence weight $2$. Each subcase contributes $q-1$ vectors, adding another $3(q-1)$ vectors of weight $2$.
        \item When exactly one of $g_2, g_3, g_4$ is zero and the first coordinate $a \neq 0$, for each of the three subcases, the number of triples with the two nonzero parameters such that $a \neq 0$ is $(q-1)^2 - (q-1) = (q-1)(q-2)$. Hence these contribute $3(q-1)(q-2)$ vectors.
        \item When all three parameters $g_2, g_3, g_4$ are nonzero and $a = 0$. The equation $h_2 g_2 + h_3 g_3 + h_4 g_4 = 0$ with all variables nonzero has $(q-1)(q-2)$ solutions (as argued above). These vectors have nonzero coordinates $g_2, g_3, g_4$ and zero first coordinate, so weight $3$. This adds $(q-1)(q-2)$ vectors of weight $3$.
        \item When all three parameters are nonzero and $a \neq 0$. The total number of triples with all nonzero is $(q-1)^3$, and we subtract those with $a = 0$ counted above, giving $(q-1)^3 - (q-1)(q-2) = (q-1)(q^2 - 3q + 3)$ vectors of weight $4$.
    \end{itemize}
    In total, the weight distribution of $\cU$ is $(0, 6(q - 1), 4(q - 1)(q - 2), (q - 1)(q^2 - 3q + 3))$.

    With the weight distributions and the number of subspaces for each distribution, we can apply Lemma~\ref{lem:formula} to obtain the desired upper bound on $T_{\max}(q, n, 4)$.
    When $k = 4$, it holds that
    $$\sum_{r = t}^{k - 1}(-1)^{r- t}q^{\binom{r- t}{2}}\left(\qbin{k - t}{r - t} - \qbin{k - t - 1}{r - t - 1}\right) = \begin{cases}
        (q - 1)^2(q + 1), & t = 1,\\
        1 - q, & t = 2,\\
        1, & t = 3.
    \end{cases}$$

    For any $0\leq w_1, w_2, w_3, w_4\leq 1$ such that $4(q - 1)w_1 + 6(q - 1)^2w_2+4(q - 1)^3w_3 + (q-1)^4w_4 = 1$, let $x = (q - 1)w_1$, $y = (q - 1)w_2$, $z = (q - 1)w_3$, and $w = (q - 1)w_4$. Let $G$ be a generator matrix such that $w_{\bu}(G) = w_{\mathrm{wt}(\bu)}$, by Lemma \ref{lem:formula} it holds that 
    \begin{align*}
        T_{\max}(q, n, 4) \leq 1 + \frac{1}{4}&\left[(q - 1)^2(q + 1)\left(12\left(\frac{1}{1 - x} - 1\right) + 24(q-1)\left(\frac{1}{1 - y} - 1\right)\right.\right.\\ 
        &+ \left.16(q - 1)^2\left(\frac{1}{1 - z} - 1\right) + 4(q - 1)^3\left(\frac{1}{1 - w} - 1\right)\right)\\
        & + (1 - q)\left(12\left(\frac{1}{1 - 2x - (q - 1)y} - 1\right) + 36(q - 1)\left(\frac{1}{1 - x - y - (q - 1)z} - 1\right)\right.\\
        &+ 16(q - 1)^2\left(\frac{1}{1 - 3y - (q - 2)z} - 1\right) + 12(q - 1)^2\left(\frac{1}{1 - x - z - (q - 1)w} - 1\right)\\
        &+ 12(q - 1)^2\left(\frac{1}{1 - 2y - (q - 1)w} - 1\right) + 24(q - 1)^3\left(\frac{1}{1 - y - 2z - (q - 2)w} - 1\right)\\
        &\left.+ 4(q - 1)^3(q - 2)\left(\frac{1}{1 - 4z - (q - 3)w} - 1\right)\right) \\
        &+\left(4\left(\frac{1}{1 - 3x - 3(q - 1)y - (q - 1)^2z} - 1\right)\right.\\
        &+ 12(q - 1)\left(\frac{1}{1 - 2x - qy - 2(q - 1)z - (q - 1)^2w} - 1\right)\\
        &+ 12(q - 1)^2\left(\frac{1}{1 - x - 3y - (4q - 5)z - (q - 1)(q - 2)w} - 1\right)\\
        &+\left.\left.4(q - 1)^3\left(\frac{1}{1 - 6y - 4(q - 2)z - (q^2 - 3q + 3)w} - 1\right)\right)\right].
    \end{align*}
\end{IEEEproof}

Now we are ready to prove Theorem~\ref{thm:k=4}. 

\begin{IEEEproof}[Proof of Theorem~\ref{thm:k=4}] Let $y = \frac{\lambda}{q - 1}$, $z = \frac{\mu}{(q - 1)^2}$, and $w = \frac{\nu}{(q - 1)^3}$, where $\lambda$, $\mu$, and $\nu$ are constants to be determined later. It then follows that $x = \frac{1 - 6\lambda - 4\mu - \nu}{4}$. In the asymptotic regime where $q \to \infty$, we first analyze each term appearing in the formula of Lemma~\ref{lem:weight distribution 4} separately. We have
\begin{align*}
    12(q - 1)^2(q + 1)\left(\frac{1}{1 - x} - 1\right) = \frac{12x}{1 - x}(q - 1)^3 + \frac{24x}{1 - x}(q - 1)^2.
\end{align*}
\begin{align*}
    24(q - 1)^3(q + 1)\left(\frac{1}{1 - y} - 1\right) 
	= &24(q - 1)^3(q + 1)(\frac{\lambda}{q - 1} + \frac{\lambda^2}{(q - 1)^2} + \frac{\lambda^3}{(q - 1)^3} + \frac{\lambda^4}{(q - 1)^4} + o(\frac{1}{q^4}))\\
    = &24\lambda(q - 1)^3 + (24\lambda^2+48\lambda)(q - 1)^2 + (24\lambda^3+48\lambda^2)(q - 1) + (24\lambda^4+48\lambda^3) + o(1).
\end{align*}
\begin{align*}
    16(q - 1)^4(q + 1)\left(\frac{1}{1 - z} - 1\right) = &16((q - 1)^5 + 2(q - 1)^4)\left(\frac{\mu}{(q - 1)^2} + \frac{\mu^2}{(q - 1)^4} + o(\frac{1}{q^5})\right)\\
    = &16\mu(q - 1)^3+32\mu(q - 1)^2 + 16\mu^2(q - 1) + 32\mu^2 + o(1).
\end{align*}
\begin{align*}
    4(q - 1)^5(q + 1)\left(\frac{1}{1 - w} - 1\right) = &4((q - 1)^6 + 2(q - 1)^5)\left(\frac{\nu}{(q - 1)^3} + \frac{\nu^2}{(q - 1)^6} + o(\frac{1}{q^6})\right)\\
    = &4\nu(q - 1)^3 + 8\nu(q - 1)^2 + 4\nu^2 + o(1).
\end{align*}
\begin{align*}
    -12(q - 1)\left(\frac{1}{1 - 2x - (q - 1)y} - 1\right) = &-\frac{24x + 12\lambda}{1 - 2x - \lambda}(q - 1).
\end{align*}
\begin{align*}
    -36(q - 1)^2\left(\frac{1}{1 - x - y - (q - 1)z} - 1\right) =& -36(q - 1)^2\left(\frac{1}{1-x}\frac{1}{1 - \frac{\lambda + \mu}{(1 - x)(q - 1)}} - 1\right)\\
    = &-36(q - 1)^2\left(\frac{x}{1-x} + \frac{\lambda+\mu}{(q - 1)(1 - x)^2} + \frac{(\lambda + \mu)^2}{(1 - x)^3(q - 1)^2} + o(\frac{1}{q^2})\right)\\
    = &-\frac{36x}{1-x}(q - 1)^2 - \frac{36(\lambda + \mu)}{(1 - x)^2}(q - 1) - \frac{36(\lambda + \mu)^2}{(1 - x)^3} + o(1).
\end{align*}
\begin{align*}
    &-16(q - 1)^3\left(\frac{1}{1 - 3y - (q - 2)z} - 1\right) \\
	= &-16(q - 1)^3\left(\frac{1}{1-\left(\frac{3\lambda + \mu}{q - 1} - \frac{\mu}{(q - 1)^2}\right)} - 1\right)\\
    =&-16(q - 1)^3\left(\frac{3\lambda + \mu}{q - 1}-\frac{\mu}{(q - 1)^2} + \frac{(3\lambda + \mu)^2}{(q - 1)^2}-\frac{2\mu(3\lambda + \mu)}{(q - 1)^3} + \frac{(3\lambda + \mu)^3}{(q - 1)^3} + o(\frac{1}{q^3})\right)\\
    = &-16(3\lambda+\mu)(q - 1)^2 - 16\left((3\lambda+\mu)^2-\mu\right)(q - 1)-16\left((3\lambda + \mu)^3-2\mu(3\lambda + \mu)\right) + o(1).
\end{align*}
\begin{align*}
    -12(q - 1)^3\left(\frac{1}{1 - x - z - (q - 1)w} - 1\right) =&-12(q - 1)^3\left(\frac{1}{1 - x}\frac{1}{1-\frac{\mu+\nu}{(1-x)(q-1)^2}} - 1\right)\\
    =&-12(q - 1)^3\left(\frac{x}{1 - x} + \frac{\mu+\nu}{(1-x)^2(q-1)^2} + o(\frac{1}{q^3})\right)\\
    = &-\frac{12x}{1-x}(q-1)^3-\frac{12(\mu+\nu)}{(1-x)^2}(q - 1) + o(1).
\end{align*}
\begin{align*}
    -12(q - 1)^3\left(\frac{1}{1 - 2y - (q - 1)w} - 1\right) =& -12(q - 1)^3\left(\frac{1}{1 - \frac{2\lambda}{q - 1} - \frac{\nu}{(q - 1)^2}} - 1\right)\\
    =&-12(q - 1)^3\left(\frac{2\lambda}{q - 1} + \frac{\nu}{(q - 1)^2} + \frac{4\lambda^2}{(q - 1)^2} + \frac{4\lambda\nu}{(q - 1)^3} + \frac{8\lambda^3}{(q - 1)^3} + o(\frac{1}{q^3})\right)\\
    = &-24\lambda(q - 1)^2 - 12(4\lambda^2+\nu)(q - 1) - 12(4\lambda\nu + 8\lambda^3) + o(1).
\end{align*}
\begin{align*}
    &-24(q - 1)^4\left(\frac{1}{1 - y - 2z - (q - 2)w} - 1\right)\\ 
	= &-24(q - 1)^4\left(\frac{1}{1 - \left(\frac{\lambda}{q - 1} + \frac{2\mu+\nu}{(q - 1)^2} - \frac{\nu}{(q - 1)^3}\right)} - 1\right)\\
    =&-24(q - 1)^4\left(\frac{\lambda}{q - 1} + \frac{2\mu+\nu +\lambda^2}{(q - 1)^2} + \frac{2\lambda(2\mu+\nu) + \lambda^3-\nu}{(q - 1)^3} + \frac{(2\mu+\nu)^2+3\lambda^2(2\mu+\nu)+\lambda^4-2\lambda\nu}{(q - 1)^4} + o(\frac{1}{q^4})\right)\\
    = & -24\lambda(q - 1)^3 -24(2\mu+\nu +\lambda^2)(q - 1)^2 - 24(2\lambda(2\mu+\nu) + \lambda^3-\nu)(q - 1)\\
    &- 24((2\mu+\nu)^2+3\lambda^2(2\mu+\nu)+\lambda^4-2\lambda\nu) + o(1).
\end{align*}
\begin{align*}
    &-4(q - 1)^4(q - 2)\left(\frac{1}{1 - 4z - (q - 3)w} - 1\right)\\=& -4\left((q - 1)^5 - (q - 1)^4\right)\left(\frac{1}{1-\left(\frac{4\mu+\nu}{(q - 1)^2} - \frac{2\nu}{(q - 1)^3}\right)} - 1\right)\\
    =&-4\left((q - 1)^5 - (q - 1)^4\right)\left(\frac{4\mu+\nu}{(q - 1)^2} - \frac{2\nu}{(q - 1)^3} + \frac{(4\mu+\nu)^2}{(q - 1)^4} - \frac{4\nu(4\mu+\nu)}{(q - 1)^5} + o(\frac{1}{q^5})\right)\\
    =& -4(4\mu+\nu)(q - 1)^3 + (8\nu+4(4\mu+\nu))(q - 1)^2 -(4(4\mu + \nu)^2 + 8\nu)(q - 1) + 16\nu(4\mu+\nu) + 4(4\mu+\nu)^2+o(1).
\end{align*}
\begin{align*}
    4\left(\frac{1}{1 - 3x - 3(q - 1)y - (q - 1)^2z} - 1\right) = \frac{4(3x+3\lambda+\mu)}{1-3x-3\lambda-\mu}
\end{align*}
\begin{align*}
    12(q - 1)\left(\frac{1}{1 - 2x - qy - 2(q - 1)z - (q - 1)^2w} - 1\right) = &12(q - 1)\left(\frac{1}{1 - 2x - \lambda - \frac{\lambda + 2\mu + \nu}{q - 1}} - 1\right)\\
    = &12(q - 1)\left(\frac{2x + \lambda}{1 - 2x - \lambda} + \frac{\lambda + 2\mu + \nu}{(1-2x-\lambda)^2(q - 1)} + o(\frac{1}{q})\right)\\
    = &\frac{12(2x+\lambda)}{1-2x-\lambda}(q - 1) + \frac{12(\lambda + 2\mu + \nu)}{(1-2x-\lambda)^2}.
\end{align*}
\begin{align*}
    &12(q - 1)^2\left(\frac{1}{1 - x - 3y - (4q - 5)z - (q - 1)(q - 2)w} - 1\right) \\
    = &12(q - 1)^2\left(\frac{1}{1-x-\left(\frac{3\lambda+4\mu+\nu}{q - 1} - \frac{\mu + \nu}{(q - 1)^2}\right)} - 1\right)\\
    = &12(q - 1)^2\left(\frac{x}{1-x} + \frac{3\lambda+4\mu+\nu}{(q - 1)(1-x)^2} - \frac{\mu+\nu}{(q - 1)^2(1 - x)^2} + \frac{(3\lambda+4\mu+\nu)^2}{(q - 1)^2(1 - x)^3} + o(\frac{1}{q^2})\right)\\
    = &\frac{12x}{1-x}(q - 1)^2 + \frac{12(3\lambda+4\mu+\nu)}{(1 - x)^2}(q - 1) + \frac{12(3\lambda + 4\mu + \nu)^2}{(1 - x)^3} - \frac{12(\mu + \nu)}{(1 - x)^2}.
\end{align*}
{\small\begin{align*}
    &4(q - 1)^3\left(\frac{1}{1 - 6y - 4(q - 2)z - (q^2 - 3q + 3)w} - 1\right) \\
    = &4(q - 1)^3\left(\frac{1}{1 - \left(\frac{6\lambda+4\mu+\nu}{q - 1} - \frac{4\mu + \nu}{(q - 1)^2} + \frac{\nu}{(q - 1)^3}\right)} - 1\right)\\
    = &4(q - 1)^3\left(\frac{6\lambda+4\mu+\nu}{q - 1} - \frac{4\mu + \nu}{(q - 1)^2} + \frac{\nu}{(q - 1)^3} + \frac{(6\lambda+4\mu+\nu)^2}{(q - 1)^2} - \frac{2(6\lambda+4\mu+\nu)(4\mu + \nu)}{(q - 1)^3} + \frac{(6\lambda+4\mu+\nu)^3}{(q - 1)^3} + o(\frac{1}{q^3})\right)\\
    = &4(6\lambda+4\mu+\nu)(q - 1)^2 + 4((6\lambda+4\mu+\nu)^2 - (4\mu + \nu))(q - 1) + 4((6\lambda+4\mu+\nu)^3 - 2(6\lambda+4\mu+\nu)(4\mu + \nu) + \nu) + o(1).
\end{align*}}

Sum up all the terms, we have the coefficients of $(q - 1)^3$, $(q - 1)^2$ and $(q - 1)$ are zero and obtain the main term as a constant independent of $q$:
\begin{align*}
    &24\lambda^4+48\lambda^3 + 32\mu^2 + 4\nu^2  - \frac{36(\lambda + \mu)^2}{(1 - x)^3}-16\left((3\lambda + \mu)^3-2\mu(3\lambda + \mu)\right) - 12(4\lambda\nu + 8\lambda^3) \\
    - &24((2\mu+\nu)^2+3\lambda^2(2\mu+\nu)+\lambda^4-2\lambda\nu) + 16\nu(4\mu+\nu) + 4(4\mu+\nu)^2+\frac{4(3x+3\lambda+\mu)}{1-3x-3\lambda-\mu} \\+ &\frac{12(\lambda + 2\mu + \nu)}{(1-2x-\lambda)^2} + \frac{12(3\lambda + 4\mu + \nu)^2}{(1 - x)^3} - \frac{12(\mu + \nu)}{(1 - x)^2}+ 4((6\lambda+4\mu+\nu)^3 - 2(6\lambda+4\mu+\nu)(4\mu + \nu) + \nu)\\
    =& 384 \lambda^3+1152 \lambda^2 \mu+360 \lambda^2 \nu+1008 \lambda \mu^2+576 \lambda \mu \nu+72 \lambda \nu^2-96 \lambda \mu-48 \lambda \nu+240 \mu^3+192 \mu^2 \nu+48 \mu \nu^2-96 \mu^2\\
    -&64 \mu \nu+4 \nu^3
    -8 \nu^2+4 \nu + \frac{-36(\lambda+\mu)^2+12(3 \lambda+4 \mu+\nu)^2}{(1-x)^3} -\frac{12(\mu+\nu)}{(1-x)^2}+\frac{4(3 x+3 \lambda+\mu)}{1-3 x-3 \lambda-\mu}+\frac{12(\lambda+2 \mu+\nu)}{(1-2 x-\lambda)^2}. 
\end{align*}

When $x\approx0.109707, \lambda\approx0.078152, \mu\approx0.012408$, and $\nu\approx0.042625$, we have $$\limsup_{q\to\infty}T_{\mathrm{ave}}(q, k = 4)\leq \limsup_{q\to\infty}T_{\max}(q, k = 4)<0.862882 \cdot 4.$$
\end{IEEEproof}

\bibliographystyle{IEEEtran}
\bibliography{ref}

\end{document}